\documentclass{article}
\usepackage{ijcai19}

\usepackage{times}
\usepackage{soul}
\usepackage{url}
\usepackage[hidelinks]{hyperref}
\usepackage[utf8]{inputenc}
\usepackage[small]{caption}
\usepackage{graphicx}
\usepackage{amsmath}
\usepackage{booktabs}
\usepackage{algorithm}
\usepackage{algorithmic}
\usepackage{xcolor}
\usepackage{soul}
\usepackage{amssymb}
\usepackage{cases}
\usepackage{comment}
\usepackage{latexsym}
\usepackage{multirow}
\usepackage{subfig}
\newtheorem{theorem}{Theorem}

\newtheorem{proof}{Proof}
\usepackage{array}
\DeclareMathOperator*{\argmax}{arg\,max}

\title{Towards Efficient Detection and Optimal Response against Sophisticated Opponents}

\author{
Tianpei Yang$^1$
\and Jianye Hao$^{1}$\footnote{Corresponding Author}\and
Zhaopeng Meng$^{1}$\and
Chongjie Zhang$^2$ \and
Yan Zheng$^1$ \and
Ze Zheng$^3$ 
\affiliations
$^1$College of Intelligence and Computing, Tianjin University\\
$^2$MMW, Tsinghua University, China\\
$^3$Beifang Investigation, Design \& Research CO.LTD
\emails
\{tpyang, jianye.hao, mengzp, yanzheng\}@tju.edu.cn,
chongjie@tsinghua.edu.cn,
zezheng0618@gmail.com
}

\begin{document}

\maketitle

\begin{abstract}  
Multiagent algorithms often aim to accurately predict the behaviors of other agents and find a best response accordingly. Previous works usually assume an opponent uses a stationary strategy or randomly switches among several stationary ones. However, an opponent may exhibit more sophisticated behaviors by adopting more advanced reasoning strategies, e.g., using a Bayesian reasoning strategy. This paper proposes a novel approach called Bayes-ToMoP which can efficiently detect the strategy of opponents using either stationary or higher-level reasoning strategies. Bayes-ToMoP also supports the detection of previously unseen policies and learning a best-response policy accordingly. We provide a theoretical guarantee of the optimality on detecting the opponent's strategies. We also propose a deep version of Bayes-ToMoP by extending Bayes-ToMoP with DRL techniques. Experimental results show both Bayes-ToMoP and deep Bayes-ToMoP outperform the state-of-the-art approaches when faced with different types of opponents in two-agent competitive games.
\end{abstract}

\section{Introduction} \label{sec1}
In multiagent systems, the ideal behavior of an agent is contingent on the behaviors of coexisting agents. However, agents may exhibit different behaviors adaptively depending on the contexts they encounter. Hence, it is critical for an agent to quickly predict or recognize the behaviors of other agents, and make a best response accordingly \cite{Powers2005Learning,fernandez2010probabilistic,albrecht2018autonomous,hernandez2017survey,ytp}. 

One efficient way of recognizing the strategies of other agents is to leverage the idea of Bayesian Policy Reuse (BPR) \shortcite{rosman2016bayesian}, which was originally proposed to determine the best policy when faced with different tasks. Hernandez-Leal et al. \shortcite{hernandez2016identifying} proposed BPR+ by extending BPR to multiagent learning settings to detect the dynamic changes of an opponent's strategies. BPR+ also extends BPR with the ability to learn new policies online against an opponent using previously unseen policies. However, BPR+ is designed for single-state repeated games only. Later, Bayes-Pepper \shortcite{hernandez2017towards} is proposed for stochastic games by combing BPR and Pepper \cite{Crandall12}. However, Bayes-Pepper cannot handle an opponent which uses a previously unknown policy. There are also some deep multiagent RL algorithms investigating how to learn an optimal policy by explicitly taking other agents' behaviors into account. He et al. \shortcite{DBLP:conf/icml/HeB16} proposed DRON which incorporates the opponent's observation into deep Q-network (DQN) and uses a mixture-of-experts architecture 
to handle different types of opponents. However, it cannot guarantee the optimality against each particular type of opponents. Recently, Zheng et al. \shortcite{yzdeepbpr18} proposed Deep BPR+ by extending BPR+ with DRL techniques to achieve more accurate detection and better response against different opponents. However, all these approaches assume that an opponent randomly switches its policies among a class of stationary ones. In practice, an opponent can exhibit more sophisticated behaviors by adopting more advanced reasoning strategies. In such situations, higher-level reasonings and more advanced techniques are required for an agent to beat such kinds of sophisticated opponents.

The above problems can be partially addressed by introducing the concept of Theory of Mind (ToM) \cite{baker2011bayesian,de2013much}. ToM is a kind of recursive reasoning technique \cite{hernandez2017survey,albrecht2018autonomous} describing a cognitive mechanism of explicitly attributing unobservable mental contents such as beliefs, desires, and intentions to other players. 
Previous methods often use nested beliefs and ``simulate" the reasoning processes of other agents to predict their actions \cite{gmytrasiewicz2005framework,wunder2012framework}.
However, these approaches show no adaptation to non-stationary opponents \cite{hernandez2017survey}. 
Later, De Weerd et al. \shortcite{de2013much} proposed a ToM model which enables an agent to predict the opponent's actions by building an abstract model of its opponent using recursive nested beliefs. Additionally, they adopt a confidence value to help an agent to adapt to different opponents. However, the main drawbacks of this model are: 1) it works only if an agent holds exactly one more layer of belief than its opponent; 2) it is designed for predicting the opponent's primitive actions instead of high-level strategies, resulting in slow adaptation to non-stationary opponents; 3) it shows poor performance against an opponent using previously unseen strategies.

To address the above challenges, we propose a novel algorithm, named Bayesian Theory of Mind on Policy (Bayes-ToMoP), which leverages the predictive power of BPR and recursive reasoning ability of ToM, to compete against such sophisticated opponents. In contrast to BPR which is capable of detecting non-stationary opponents only, Bayes-ToMoP incorporates ToM into BPR to quickly and accurately detect not only non-stationary, and more sophisticated opponents and compute a best response accordingly. Theoretical guarantees are provided for the optimal detection of the opponent's strategies. Besides, Bayes-ToMoP also supports detecting whether an opponent is using a previously unseen policy and learning an optimal response against it. Furthermore, Bayes-ToMoP can be straightforwardly extended to DRL environment with a neural network as the value function approximator, termed as deep Bayes-ToMoP. Experimental results show that both Bayes-ToMoP and deep Bayes-ToMoP outperform the state-of-the-art approaches when faced with different types of opponents in two-agent competitive games.

\section{Background} \label{sec3}
\textbf{Bayesian Policy Reuse}\quad BPR was originally proposed in \cite{rosman2016bayesian} as a framework for an agent to quickly determine the best policy to execute when faced with an unknown task. Given a set of previously-solved tasks $\mathcal{T}$ and an unknown task $\tau^{*}$, the agent is required to select the best policy $\pi^{*}$ from the policy library $\Pi$ within as small numbers of trials as possible. BPR uses the concept of \textit{belief} $\beta$, which is a probability distribution over the set of tasks $\mathcal{T}$, to measure the degree to which $\tau^{*}$ matches the known tasks based on the signal $\sigma$. 
A signal $\sigma$ can be any information that is correlated with the performance of a policy (e.g., immediate rewards, episodic returns). BPR involves performance models of policies on previously-solved tasks, which describes the distribution of returns from each policy $\pi$ on previously-solved tasks. A performance model $P(U|\tau, \pi)$ is a probability distribution over the utility of a policy $\pi$ on a task $\tau$.

A number of BPR variants with exploration heuristics are proposed to select the best policy, e.g., probability of improvement (BPR-PI) heuristic and expected improvement (BPR-EI) heuristic. BPR-PI heuristic utilizes the probability with which a policy can achieve a hypothesized increase in performance ($U^{+}$) over the current best estimate $\bar{U} = \max_{\pi \in \Pi}\sum_{\tau \in \mathcal{T}}\beta(\tau)E[U|\tau,\pi]$. Formally, it chooses the policy that most likely to achieve the utility $U^{+}$: 
$\pi^{*} = \argmax_{\pi \in \Pi}\sum_{\tau \in \mathcal{T}}\beta(\tau)P(U^{+}|\tau,\pi)$
. However, it is not straightforward to determine the appropriate value of $U^{+}$, thus another way of avoiding this issue is BPR-EI heuristic, which selects the policy most likely to achieve any possible utilities of improvement $\bar{U} < U^{+} < U^{max}$: 
$\pi^{*} = \argmax_{\pi \in \Pi}\int_{\bar{U}}^{U^{max}}\sum_{\tau \in \mathcal{T}}\beta(\tau)P(U^{+}|\tau, \pi)dU^{+}
$. 
BPR \shortcite{rosman2016bayesian} showed BPR-EI heuristic performs best among all BPR variants. Therefore, we choose BPR-EI heuristic for playing against different opponents.

\textbf{Theory of Mind} \quad ToM model \cite{de2013much} is used to predict an opponent's action by building an abstract model of the opponent using recursive nested beliefs. ToM model is described in the context of a two-player competitive game where an agent and its opponent differ in their abilities to make use of ToM. The notion of ToM$_{k}$ indicates an agent that has the ability to use ToM up to the $k$-th order, and we briefly introduce the first two orders of ToM models

A zero-order ToM (ToM$_{0}$) agent holds its zero-order belief in the form of a probability distribution on the action set of its opponent. The ToM$_{0}$ agent then chooses the action that maximizes its expected payoff.
A first-order ToM agent (ToM$_{1}$) keeps both zero-order belief $\beta^{(0)}$ and first-order belief $\beta^{(1)}$. The first-order belief $\beta^{(1)}$ is a probability distribution that describes what the ToM$_{1}$ agent believes its opponent believes about itself. The ToM$_{1}$ agent first predicts its opponent's action under its first-order belief. Then, the ToM$_{1}$ agent integrates its first-order prediction with the zero-order belief and uses this integrated belief in the final decision. The degree to which the prediction influences the agent’s actions is determined by its first-order confidence $0 \leq c_{1} \leq 1$, which is increased if the prediction is right while decreased otherwise.


\section{Bayes-ToMoP} \label{sec4}
\subsection{Motivation}\label{sec4.1}
Previous works \cite{hernandez2016identifying,hernandez2017towards,hernandez2017efficiently,yzdeepbpr18,DBLP:conf/icml/HeB16} assume that an opponent randomly switches its policies among a number of stationary policies. However, a more sophisticated agent may change its policy in a more principled way. For instance, it first predicts the policy of its opponent and then best responds towards the estimated policy accordingly. If the opponent's policy is estimated by simply counting the action frequencies, it is then reduced to the well-known fictitious play \cite{Shoham2009Multiagent}. However, in general, an opponent's action information may not be observable during interactions. One way of addressing this problem is using BPR, which uses Bayes' rule to predict the policy of the opponent according to the received signals (e.g., rewards), and can be regarded as the generalization of fictitious play. 

Therefore, a question naturally arises: how an agent can effectively play against both simple opponents with stationary strategies and more advanced ones (e.g., using BPR)? To address this question, we propose a new algorithm called Bayes-ToMoP, which leverages the predictive power of BPR and recursive reasoning ability of ToM to predict the strategies of such opponents and behave optimally. We also extend Bayes-ToMoP to DRL scenarios with a neural network as the value function approximator, termed as deep Bayes-ToMoP. In the following descriptions, we do not distinguish whether a policy is represented in a tabular form or a neural network unless necessary.

We assume the opponent owns a class of candidate stationary strategies to select from periodically. Bayes-ToMoP needs to observe the reward of its opponent which is not an assumption since in competitive settings, e.g., zero-sum games, an agent’s opponent’s reward is always the opposite of its own. Bayes-ToMoP does not need to observe the actions of its opponent except for learning a policy against an unknown opponent strategy. We use the notation of Bayes-ToMoP$_{k}$ to denote an agent with the ability of using Bayes-ToMoP up to the $k$-th order. Intuitively, Bayes-ToMoP$_{i}$ with a higher-order theory of mind could take advantage of any Bayes-ToMoP$_{j}$ with a lower-order one ($i>j$). De Weerd et al. \shortcite{de2013much} showed that the reasoning levels deeper than 2 do not provide additional benefits, so we focus on Bayes-ToMoP$_{0}$ and Bayes-ToMoP$_{1}$. Bayer-ToMoP$_{k}$ ($k>1$) can be naturally constructed by incorporating a higher-order ToM idea into our framework. 


\subsection{Bayes-ToMoP$_{0}$ Algorithm} \label{sec4.2}
We start with the simplest case: Bayes-ToMoP$_{0}$, which extends ToM${_0}$ by incorporating Bayesian reasoning techniques to predict the strategy of an opponent. Bayes-ToMoP$_{0}$ holds a zero-order belief $\beta^{(0)}$ about its opponent's strategies $\lbrace j | j \in \mathcal{J}\rbrace$, each of which $\beta^{(0)}(j)$ is a probability that its opponent may adopt each strategy $j$: $\beta^{(0)}(j) \geq 0 ,\forall j \in \mathcal{J} $.
$\sum_{j \in \mathcal{J}}\beta^{(0)}(j) = 1 $. Given a utility $U$, a performance model $P_{self}(U|j, \pi)$ describes the probability of an agent using a policy $\pi \in \Pi$ against an opponent's strategy $j$.


For Bayes-ToMoP$_{0}$ agent equipped with a policy library $\Pi$ against its opponent's  with a strategy library $\mathcal{J}$, it first initializes its performance models $P_{self}(U|\mathcal{J}, \Pi)$ and zero-order belief $\beta^{(0)}$. Then, in each episode, given the current belief $\beta^{(0)}$, Bayes-ToMoP$_{0}$ agent evaluates the expected improvement utility defined following BPR-EI heuristic for all policies and then selects the optimal one. Next, Bayes-ToMoP$_{0}$ agent updates its zero-order belief using Bayes' rule \cite{rosman2016bayesian}. At last, Bayes-ToMoP$_{0}$ detects whether its opponent is using a previously unseen policy. If yes, it learns a new policy against its opponent. The new strategy detection and learning algorithm will be described in Section \ref{sec4.4}.

Finally, note that without the new strategy detection and learning phase, Bayes-ToMoP$_{0}$ agent is essentially equivalent to BPR and Bayes-Pepper since they both first predict the opponent's strategy (or taks type) and then select the optimal policy following BPR heuristic, each strategy of the opponent here can be regarded as a task in the original BPR. Besides, the full Bayes-ToMoP$_{0}$ is essentially equivalent to BPR+ since both can handle previously unseen strategies.

\subsection{Bayes-ToMoP$_{1}$ Algorithm} \label{sec4.3}
Next, we move to Bayes-ToMoP$_{1}$ algorithm. Apart from its zero-order belief, Bayes-ToMoP$_{1}$ also maintains a first-order belief, which is a probability distribution that describes the probability that an agent believes his opponent believes it will choose a policy $\pi \in \Pi$. 
The overall strategy of Bayes-ToMoP$_{1}$ is shown in Algorithm \ref{tom1}. Given the policy library $\Pi$ and $\mathcal{J}$, performance models $P_{self}(U|\mathcal{J}, \Pi)$ and $P_{oppo}(U|\Pi, \mathcal{J})$, zero-order belief $\beta^{(0)}$ and first-order belief $\beta^{(1)}$, Bayes-ToMoP$_{1}$ agent first predicts the policy $\hat{j}$ of its opponent assuming the opponent maximizes its own utility based on BPR-EI heuristic under its first-order belief (Line 2). 
However, this prediction may conflict with its zero-order belief. To address this conflict, Bayes-ToMoP$_{1}$ holds a first-order confidence $c_{1}(0 \leq c_{1} \leq 1)$ serving as the weighting factor to balance the influence between its first-order prediction and zero-order belief. Then, an integration function $I$ is introduced to compute the final prediction results which is defined as the linear combination of the first-order prediction $\hat{j}$ and zero-order belief $\beta^{(0)}$ weighted by the confidence degree $c_{1}$ following Equation \ref{eq9:u} \cite{de2013much} (Line 3). 
\begin{equation} \label{eq9:u}
I(\beta^{(0)},\hat{j},c_{1})(j)=
\begin{cases}
(1-c_{1})\beta^{(0)}(j) + c_{1} & \; \text{if } j = \hat{j}\\
(1-c_{1})\beta^{(0)}(j) & \; \text{otherwise }
\end{cases}
\end{equation} 
Next, Bayes-ToMoP$_{1}$ agent computes the optimal policy based on the integrated belief (Line 4).
At last, Bayes-ToMoP$_{1}$ updates its first-order belief and zero-order belief using Bayes' rule \cite{rosman2016bayesian} (Lines 6-11).   

\begin{algorithm}[h] 
\caption{Bayes-ToMoP$_{1}$ Algorithm} \label{tom1}
\textbf{Initialize:} Policy library $\Pi$ and $\mathcal{J}$, performance models $P_{self}(U|\mathcal{J}, \Pi)$ and $P_{oppo}(U|\Pi, \mathcal{J})$, zero-order belief $\beta^{(0)}$, first-order belief $\beta^{(1)}$
\begin{algorithmic}[1]
\FOR{each episode}
\STATE Compute the first-order opponent policy prediction 
$\hat{j}$:\\ $\argmax_{j \in \mathcal{J}}\int_{\bar{U}}^{U^{max}}\sum_{\pi \in \Pi}\beta^{(1)}(\pi)P_{oppo}(U^{+}|\pi, j)dU^{+}$\\
\STATE Integrate $\hat{j}$ with $\beta^{(0)}$: $I(\beta^{(0)},\hat{j},c_{1})$ (see Equation (\ref{eq9:u}))
\STATE Select the optimal policy $\pi^{*}$: \\ 
$\argmax \limits_{\pi \in \Pi}\int_{\bar{U}}^{U^{max}}\sum \limits_{j' \in \mathcal{J}}I(\beta^{(0)},\hat{j},c_{1})(j)P_{self}(U^{+}|j, \pi)dU^{+}$ \\
\STATE Play and receive the episodic return $\langle r_{self},r_{oppo}\rangle$
\FOR{each own policy $\pi \in \Pi$}
\STATE Update first-order belief $\beta^{(1)}$:\\
$\beta^{(1)}(\pi)= 
\frac{P_{oppo}(r_{oppo}|\pi, \hat{j})\beta^{(1)}(\pi)}{\sum_{\pi^{'}\in \Pi}P_{oppo}(r_{oppo}|\pi^{'}, \hat{j})\beta^{(1)}(\pi^{'})}
$
\ENDFOR
\FOR{each opponent strategy $j \in \mathcal{J}$}
\STATE Update zero-order belief $\beta^{(0)}$:\\
$\beta^{(0)}(j)= 
\frac{P_{self}(r_{self}|j, \pi)\beta^{(0)}(j)}{\sum_{j^{'}\in \mathcal{J}}P_{self}(r_{self}|j^{'}, \pi)\beta^{(0)}(j^{'})}
$
\ENDFOR
\STATE Update $c_{1}$ following Equation (\ref{eq12:c})
\STATE Detect new opponent strategy
\ENDFOR
\end{algorithmic}
\end{algorithm}

The next issue is how to update the first-order confidence degree $c_{1}$. The value of $c_{1}$ can be understood as the exploration rate of using first-order belief to predict the opponent's strategies. In previous ToM model \cite{de2013much}, the value of $c_{1}$ is increased if the prediction is right while decreased otherwise based on the assumption that an agent can observe the actions of its opponent. 
However, in our settings, the prediction works on a higher level of behaviors (the policies), which information usually is not available (agents are not willing to reveal their policies to others to avoid being exploited in competitive environments). Therefore, we propose to use game outcomes as the signal to indicate whether our previous predictions are correct and adjust the first-order confidence degree accordingly. Specifically, in a competitive environment, we can distinguish game outcomes into three cases: win ($r_{self} > r_{oppo}$), lose or draw. Thus, the value of $c_{1}$ is increased when the agent wins and decreased otherwise by an adjustment rate of $\lambda$. Following this heuristic, Bayes-ToMoP$_{1}$ can easily take advantage of Bayes-ToMoP$_{0}$ since it can well predict the policy of Bayes-ToMoP$_{0}$ in advance. However, this heuristic does not work with less sophisticated opponents, e.g., an opponent simply switching among several stationary policies without the ability of using ToM. 
This is due to the fact that the curve of $c_{1}$ becomes oscillating when it is faced with an agent who is unable to make use of ToM, thus fails to predict the opponent's behaviors accurately. 

To this end, we propose an adaptive and generalized mechanism to adjust the value of $c_{1}$. We first introduce the concept of win rate $\upsilon_{i}=\frac{\sum_{i-l}^{i}r_{self}}{l}$ during a fixed length $l$ of episodes. Since Bayes-ToMoP$_{1}$ agent assumes its opponent is Bayes-ToMoP$_{0}$ at first, the value of $l$ controls the number of episodes before considering its opponent may switch to a less sophisticated type. If the average performance till the current episode is better than the previous episode's ($\upsilon_{i}\geq \upsilon_{i-1}$), we increase the weight of using first-order prediction, i.e., increasing the value of $c_{1}$ with an adjustment rate $\lambda$; if $\upsilon_{i}$ is less than $\upsilon_{i-1}$ but still higher than a threshold $\delta$, it indicates the performance of the first-order prediction diminishes. Then Bayes-ToMoP$_{1}$ decreases the value of $c_{1}$ quickly with a decreasing factor $\frac{lg\upsilon_{i}}{lg(\upsilon_{i} - \delta)}$; if $\upsilon_{i} \leq \delta$, the rate of exploring first-order belief is set to 0 and only zero-order belief is used for prediction. Formally we have:
\begin{equation} \label{eq12:c}
c_{1}=
\begin{cases}
((1-\lambda)c_{1} + \lambda)\mathbf{F}(\upsilon_{i}) & \quad \text{if }  \upsilon_{i} \geq \upsilon_{i-1} \\
(\frac{lg\upsilon_{i}}{lg(\upsilon_{i} - \delta)}c_{1})\mathbf{F}(\upsilon_{i}) & \quad \text{if } \delta < \upsilon_{i} < \upsilon_{i-1} \\
\lambda \mathbf{F}(\upsilon_{i}) & \quad \text{if } \upsilon_{i} \leq \delta
\end{cases}
\end{equation}
where $\delta$ is the threshold of the win rate $\upsilon_{i}$, which reflects the lower bound of the difference between its prediction and its opponent's actual behaviors. $\mathbf{F}(\upsilon_{i})$ is an indicator function to control the direction of adjusting the value of $c_{1}$. Intuitively, Bayes-ToMoP$_{1}$ detects the switching of its opponent's strategies at each episode $i$ and reverses the value of $\mathbf{F}(\upsilon_{i})$ whenever its win rate $\upsilon_{i}$ is no larger than $\delta$ (Equation \ref{eq14:o}). 
\begin{equation} \label{eq14:o}
\mathbf{F}(\upsilon_{i}):= \begin{cases}1& \text{if (} \upsilon_{i} \leq \delta \; \& \; \mathbf{F}(\upsilon_{i}) = 0 ) \\ 0&\text{if (} \upsilon_{i} \leq \delta \; \& \; \mathbf{F}(\upsilon_{i}) = 1) \end{cases}
\end{equation} 
Finally, Bayes-ToMoP$_{1}$ learns a new optimal policy if it detects a new opponent strategy (detailed in next section).
\subsection{New Opponent Detection and Learning} \label{sec4.4}
The new opponent detection and learning component is the same for all Bayes-ToMoP$_k$ agents ($k\geq0$). Bayes-ToMoP$_k$ first detects whether its opponent is using a new kind of strategies. This is achieved by recording a fixed length of game outcomes and checking whether its opponent is using one of the known strategies at each episode. In details, Bayes-ToMoP$_k$ keeps a length $h$ of memory recording the game outcomes at episode $i$, and uses the win rate $\theta_{i}=\frac{\sum_{i-h}^{i}r_{self}}{h}$ over the most recent $h$ episodes as the signal indicating the average performance over all policies till the current episode $i$. If the win rate $\theta_{i}$ is lower than a given threshold $\delta$ ($\theta_{i}<\delta$), it indicates that all existing policies show poor performance against the current opponent strategy, 
in this way Bayes-ToMoP$_k$ agent infers that the opponent is using a previously unseen policy outside the current policy library. 


Since we can easily obtain the average win rate $\theta^{(\pi j)}$ of each policy $\pi$ against each known opponent strategy $j$, the lowest win rate among the best-response policies ($\min_{\pi \in \Pi} \max_{j \in \mathcal{J}} \theta^{(\pi j)}$) can be seen as an upper bound of the value of $\delta$. The value of $h$ controls the number of episodes before considering whether the opponent is using a previously unseen strategy. Note that the accuracy of detection is increased with the increase of the memory length $h$, however, a larger value of $h$ would necessarily increase the detection delay. The optimal memory length is determined empirically through extensive simulations (see supplementary materials). 

After detecting the opponent is using a new strategy, the agent begins to learn the best-response policy against it. Following previous work \cite{hernandez2016identifying}, we adopt the same assumption that the opponent will not change its strategy during the learning phase (a number of rounds). Otherwise, the learning process may not converge. For tabular Bayes-ToMoP, we adopt the traditional model-based RL: R-max \cite{BrafmanT01} to compute the optimal policy. Specifically, once a new strategy is detected, R-max estimates the state transition function $T$ and reward function $R$ with $\hat{T}$ and $\hat{R}$. It also keeps a transition count $c(s,a,s^{'})$ and total reward $t(s,a)$ for all state-action pairs. Each transition $\langle s,a,s^{'},r \rangle$ results in an update for the transition count: $c(s,a,s^{'})\gets c(s,a,s^{'}) + 1$ and total reward: $t(s,a)\gets t(s,a) + r$. The estimates $\hat{T}$ and $\hat{R}$ is updated using $c(s,a,s^{'})$ and $t(s,a)$ with a given parameter $n$: $\hat{T}(s,a,s^{'}) = c(s,a,s^{'}) / n$, $\hat{R}(s,a) = t(s,a) / n$ if $\sum_{s^{'}}c(s,a,s^{'}) \geq n$. R-max computes $\hat{Q}(s,a)=\hat{R}(s,a)+\gamma\sum_{s^{'}}\hat{T}(s,a,s^{'})\max_{a^{'}}\hat{Q}(s^{'},a^{'})$ for all state-action pairs and selects the action that maximizes $\hat{Q}(s,a)$ according to $\epsilon$-greedy mechanism. 

For deep Bayes-ToMoP, we apply DQN \cite{mnih2015human} to do off-policy learning using the obtained interaction experience. DQN is a deep Q-learning method with experience replay, consisting of a neural network approximating $Q(s,a;\theta)$ and a target network approximating $Q(s,a;\theta^{-})$. 
DQN draws samples (or minibatches) of experience $(s,a,s^{'},r) \sim U(\mathcal{D})$ uniformly from a replay memory $\mathcal{D}$, and updates using the following loss function: $L(\theta)=\mathbf{E}_{(s,a,s^{'},r),r\sim U(\mathcal{D})}[(r+\gamma \max_{a'}Q(s',a';\theta^{-})-Q(s,a;\theta))^2] $. DQN is not a requirement, actually, our learning framework is general in which other DRL approaches can be applied as well. However, most DRL algorithms suffer from the sample efficiency problems under some specific settings. This can be addressed by incorporating sample efficient DRL methods in the future. To generate new performance models, we use a neural network to estimate the policy of the opponent based on the observed state-action history of the opponent using supervised learning techniques \cite{yzdeepbpr18,foerster2018learning}. 

After the above learning phase, new performance models are generated using rewards obtained from a number of simulations of the agent’s policy against the opponent’s estimated strategy. These values are modeled as a Gaussian distribution to obtain the performance models. 
Finally, it adds the new policy $\pi_{new}$ and the estimated opponent policy to its policy library $\Pi$ and its opponent's policy library $\mathcal{J}$ respectively.

\subsection{Theoretical Analysis}\label{sec4.5}
In this section, we provide a theoretical analysis that Bayes-ToMoP can accurately detect the opponent's strategy from a known policy library and derives an optimal response policy accordingly. 

\begin{theorem} \label{theorem1}
\textbf{(Optimality on Strategy Detection)} If the opponent plays a strategy from the known policy library, Bayes-ToMoP can detect the strategy w.p.1 and selects an optimal response policy accordingly.
\end{theorem}

The proof is given in supplementary materials.

\section{Simulations} \label{sec5}
In this section, we present experimental results of Bayes-ToMoP compared with state-of-the-art tabular approaches (BPR+ \cite{hernandez2016identifying} and Bayes-Pepper \cite{hernandez2017towards}). For deep Bayes-ToMoP, we compare with the following four baseline strategies: 1) BPR$+$, 2) Bayes-Pepper (BPR+ and Bayes-Pepper use a neural network as the value function approximator), 3) DRON \cite{DBLP:conf/icml/HeB16} and 4) deep BPR+ \cite{yzdeepbpr18}. We first evaluate the performance of Bayes-ToMoP by comparing it with state-of-the-art approaches in both tabular and deep settings. We also compare the performance of Bayes-ToMoP and deep Bayes-ToMoP with previous works against an opponent using previously unseen strategies. 
The network structure and details of parameter settings are in supplementary materials.

\subsection{Game Settings} \label{sec5.1}
\begin{figure}
	\begin{minipage}[t]{.49\linewidth} 
	\centerline{\includegraphics[height = 1in]{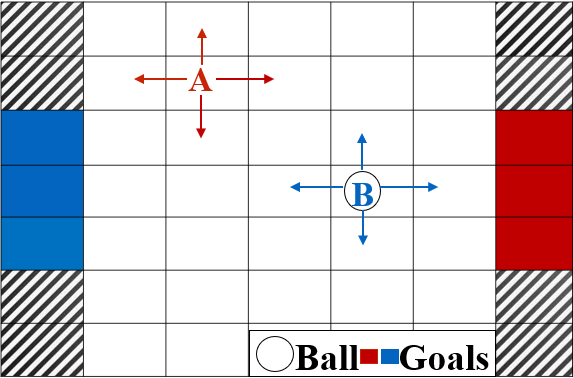}}
	\caption{The soccer game.} \label{figure1}
	\end{minipage} 
	\begin{minipage}[t]{.49\linewidth} 
	\centerline{\includegraphics[height = 1in]{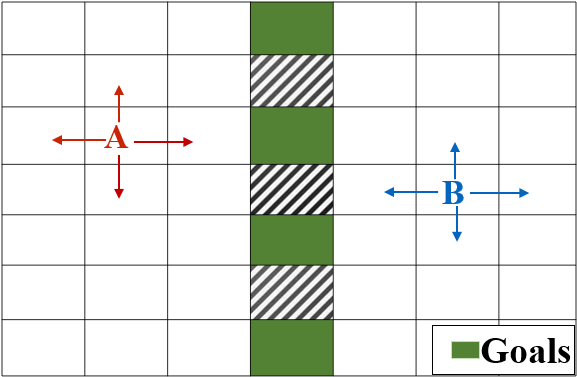}}
	\caption{Thieves and hunters} \label{figure2}
	\end{minipage}
\end{figure}

We evaluate the performance of Bayes-ToMoP on the following testbeds: 
soccer \cite{DBLP:conf/icml/Littman94,DBLP:conf/icml/HeB16} and thieves and hunters \cite{Goodrich2003Neglect,Crandall12}. 
Soccer (Figure \ref{figure1}) and Thieves and hunters (Figure \ref{figure2}) are two stochastic games both on a $7\times7$ grid. Two players, A and B, start at one of starting points in the left and right respectively and can choose one of the following 5 actions: go left, right, up, down and stay. Any action that goes to grey-slash grids or beyond the border is invalid. 1) In soccer, the ball (circle) is randomly assigned to one player initially. The possession of the ball switches when two players move to the same grid. A player scores one point if it takes the ball to its opponent's goals; 2) in thieves and hunters, player A scores one point if two players move to one goal simultaneously (A catches B), otherwise, it loses one point if player B moves to one goal without being caught. If neither player gets a score within 50 steps, the game ends with a tie. 

We consider two versions of the above games in both tabular and deep representations. For the tabular version, the state space includes few numbers of discrete states, in which Q-values can be represented in a tabular form. For the deep version, each state consists of different dimensions of information: for example, states in soccer includes coordinates of two agents and the ball possession. In this case, we evaluate the performance of deep Bayes-ToMoP. We manually design six kinds of policies for the opponent in soccer (differentiated by the directions of approaching the goal) and twenty-four kinds of policies for the opponent in thieves and hunters (differentiated by the orders of achieving the goal). A policy library of best-response policies are trained using Q-learning and DQN for Bayes-ToMoP and deep Bayes-ToMoP. 

\begin{table}[H]
	\small
\centering
	 \caption{Average win rates with std.dev.($\pm$) in soccer.}\label{table2} 	
	\begin{tabular}{cccc}
	\hline
	\multirow{2}{1.6cm}{\centering Opponents /\\ Methods} & \multirow{2}{0.6cm}{\centering O$_{ToMoP_{0}}$}& \multirow{2}{0.5cm}{O$_{ns}$} & \multirow{2}{1.6cm}{O$_{ToMoP_{0}}$-$s$}\\
     & \\
	\hline
	BPR$_s$  & 49.78$\pm$1.71\% & 99.37$\pm$0.72\% & 66.31$\pm$0.57\%  \\
	
	DRON & 74.75$\pm$0.19\% & 76.54$\pm$0.16\% & 75.35$\pm$0.18\% \\
	
	Deep BPR+ & 71.57$\pm$1.26\% & 99.49$\pm$0.51\% &  78.88$\pm$0.76\%\\
	
	Bayes-ToMoP$_{1}$ & \textbf{99.82$\pm$0.18\%} & 98.21$\pm$0.37\% & \textbf{98.48$\pm$0.54\%} \\
	\hline
	\end{tabular}	
\end{table}

\subsection{Performance against Different Opponents} \label{sec5.2}


Three kinds of opponents are considered: 1) a Bayes-ToMoP$_{0}$ opponent (O$_{ToMoP_{0}}$); 2) an opponent that randomly switches its policy among stationary strategies and lasts for an unknown number of episodes (O$_{ns}$) and 3) an opponent switching its strategy between stationary strategies and Bayes-ToMoP$_{0}$ (O$_{ToMoP_{0}}$-$s$). We assume an opponent only selects a policy from the known policy library. Thus Bayes-Pepper is functionally equivalent to BPR+ in our setting and we use BPR$_s$ to denote both strategies. Due to the space limitation, we only give experiments on the tabular form of Thieves and hunters and a deep version of soccer. 
 
\begin{figure}[H]
\centering
   \subfloat[Thieves and hunters]{
  \begin{minipage}[t]{.47\linewidth}    
	\centerline{\includegraphics[height = 1in]{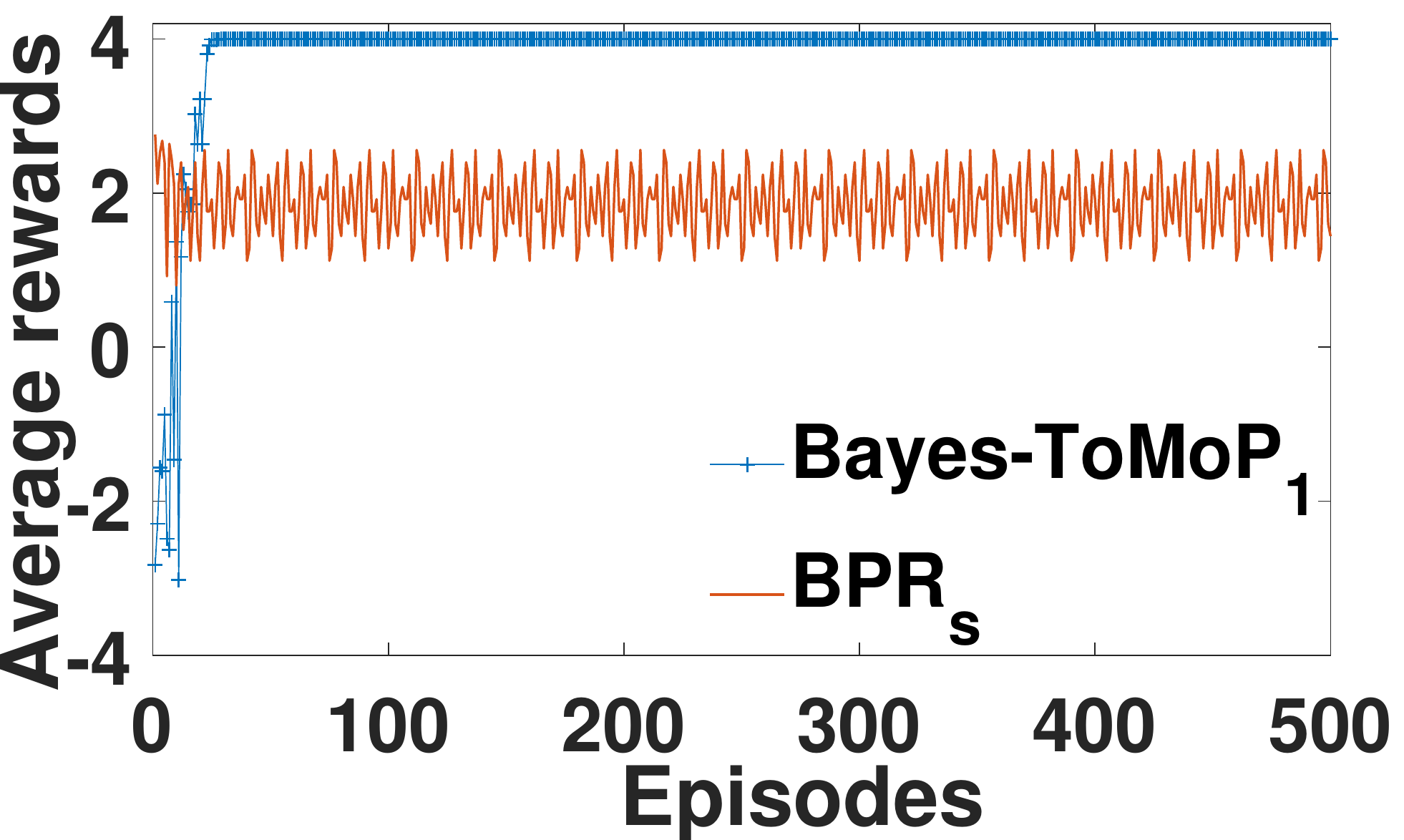}}	
  \end{minipage}
  } 
  \hfill
   \subfloat[Soccer (deep version)]{
  \begin{minipage}[t]{.47\linewidth}    
	\centerline{\includegraphics[height = 1in]{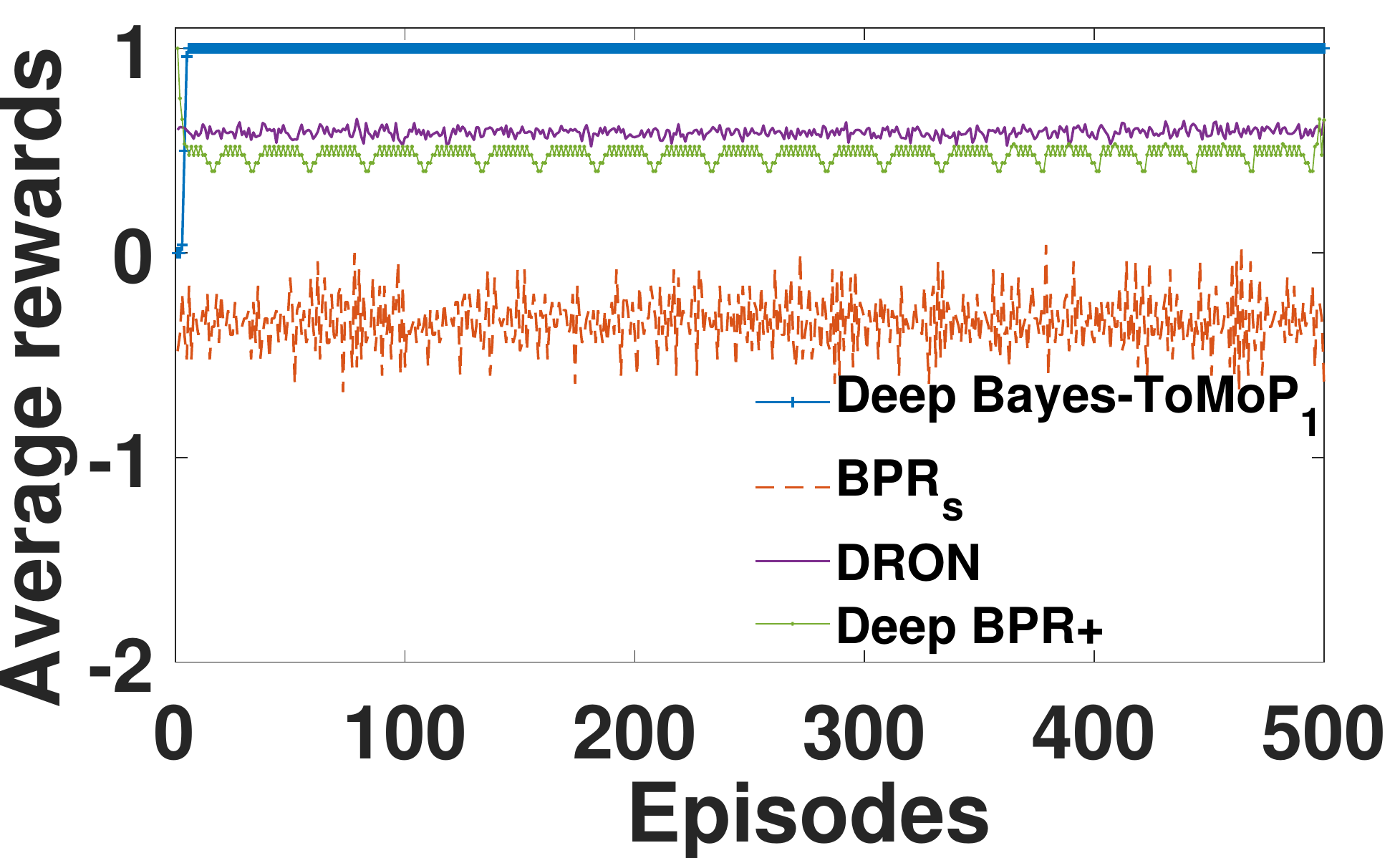}}	
  \end{minipage} 
 }
\caption{Against an opponent O$_{ToMoP_{0}}$ on different games.} \label{figure3}
\end{figure}
Figure \ref{figure3} (a) shows that only Bayes-ToMoP$_{1}$ can quickly detect the opponent's strategies and achieve the highest average rewards. In contrast, BPR$_s$ fails against O$_{ToMoP_{0}}$. Similar comparison results can be found for deep Bayes-ToMoP$_{1}$ (Figure \ref{figure3} (b)). This is because Bayes-ToMoP$_{1}$ explicitly considers two-orders of belief to do recursive reasoning first and then derives an optimal policy against its opponent. However, BPR$_s$ is essentially equivalent to Bayes-ToMoP$_{0}$, i.e., it is like Bayes-ToMoP$_{0}$ is under self-play. Therefore, neither BPR$_s$ nor Bayes-ToMoP$_{0}$ could take advantage of each other and the winning percentages are expected to be approximately 50\%. Average win rates shown in Table \ref{table2} also confirm our hypothesis. The results for Bayes-ToMoP$_{1}$ under self-play can be found in supplementary materials. Deep BPR+ incorporates previous opponent's behaviors into BPR, however, their model only considers which kind of stationary strategy the opponent is using, thus is not enough to detect the policy of opponent O$_{ToMoP_{0}}$. Figure \ref{figure3} (b) shows DRON performs better than Deep BPR+ and BPR$_s$ 
since it explicitly considers the opponent's strategies. However, it fails to achieve the highest average rewards because the dynamic adjustment of DRON cannot guarantee that the optimality against a particular type of opponents.



\begin{figure}
   \subfloat[Thieves and hunters]{
  \begin{minipage}[t]{.47\linewidth}    
	\centerline{\includegraphics[height = 1in]{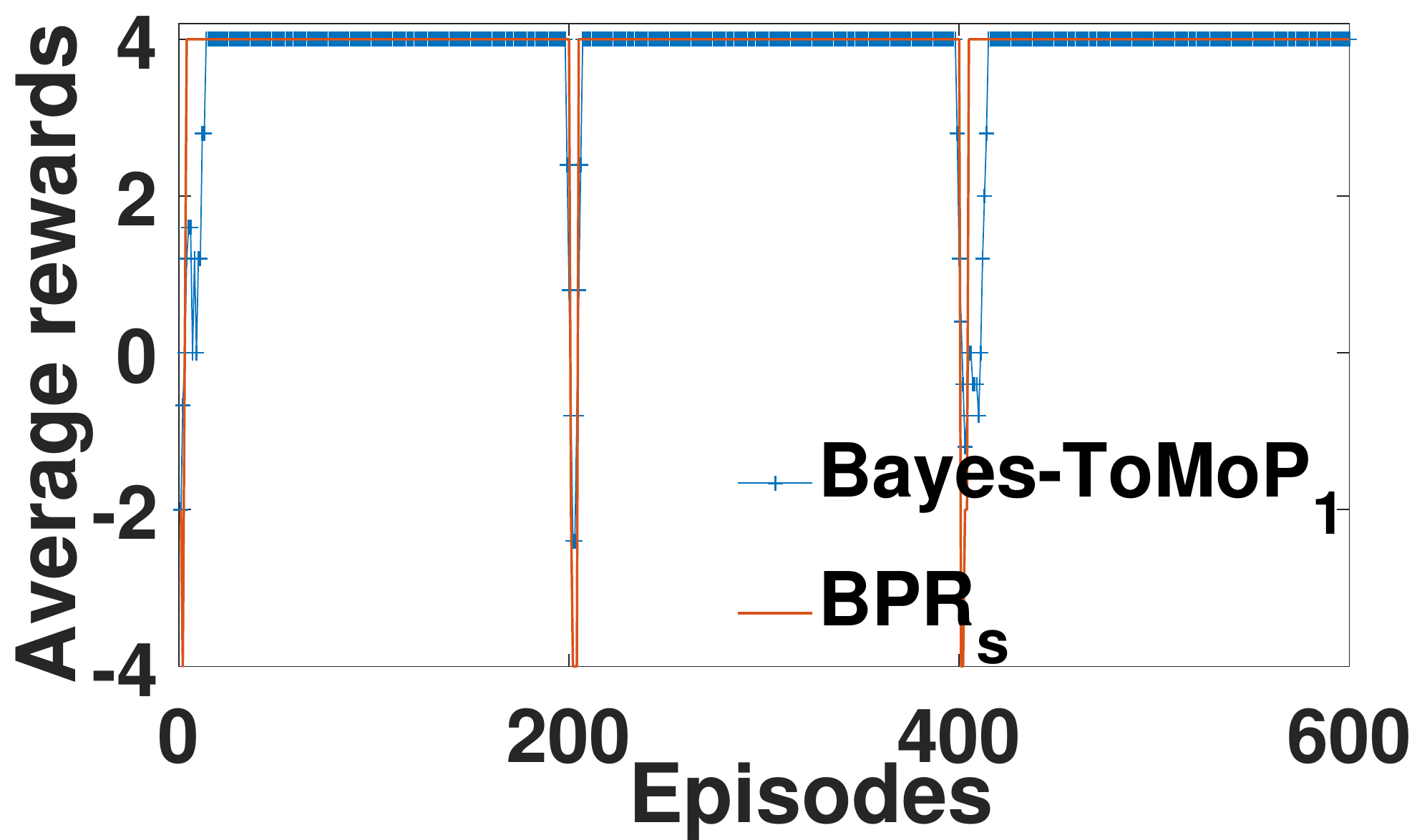}}	
  \end{minipage} 
 }
 \hfill
	\subfloat[Soccer (deep version)]{
   \begin{minipage}[t]{.47\linewidth}    
	\centerline{\includegraphics[height = 1in]{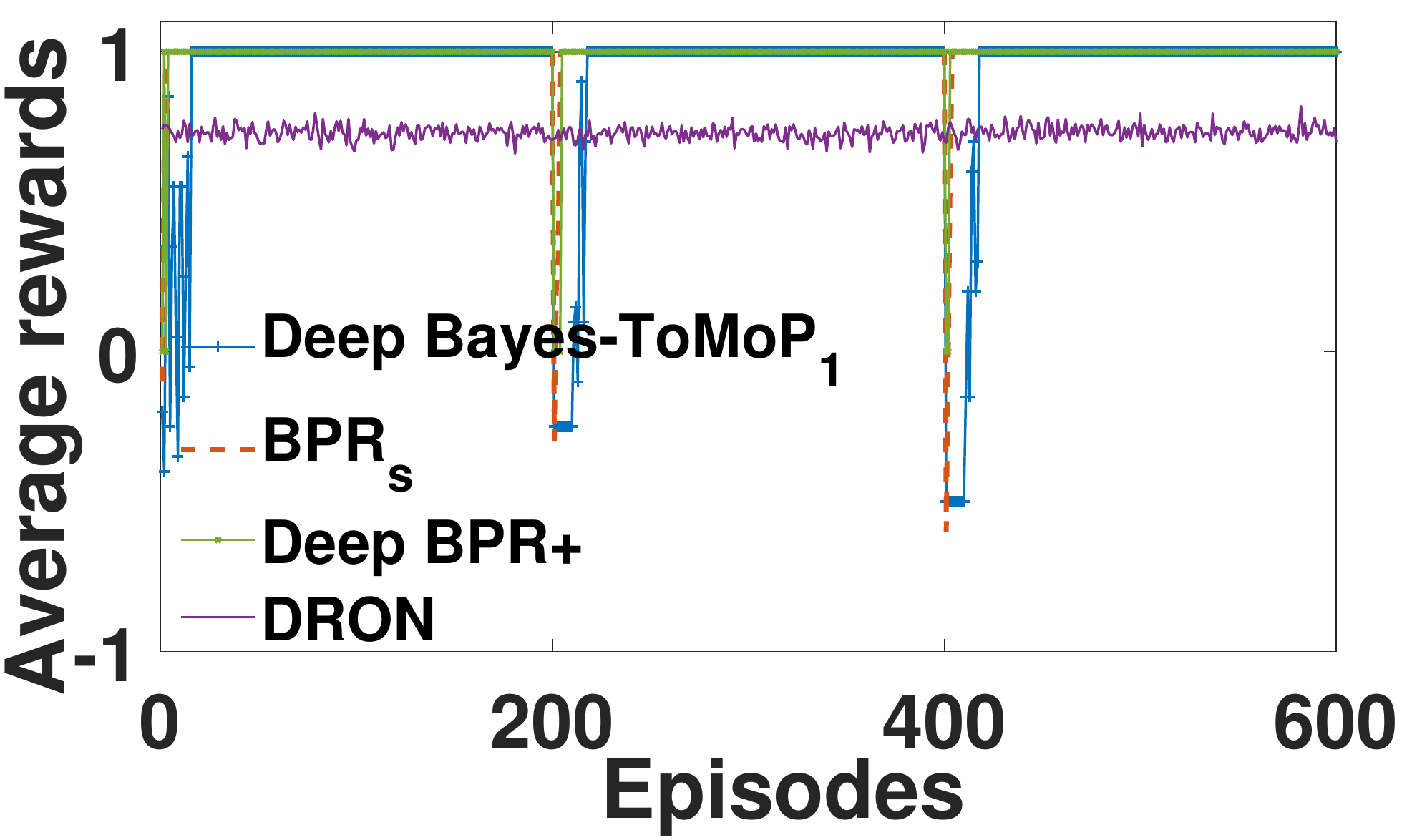}}	
  \end{minipage} 
   }
\caption{Against an opponent O$_{ns}$ on different games.} \label{figure4}
\end{figure}

Next, we present the results of playing against O$_{ns}$ that switches its policy among stationary strategies and lasts for 200 episodes. Figure \ref{figure4} (a) shows the comparison of Bayes-ToMoP$_{1}$ with BPR$_s$, where both methods can quickly and accurately detect which stationary strategy the opponent is using and derive the optimal policy against it. We observe that Bayes-ToMoP$_{1}$ requires longer time to detect than BPR$_s$ and similar comparison is found in Figure \ref{figure4} (b) that deep Bayes-ToMoP$_{1}$ requires longer time to detect than BPR$_s$ and deep BPR+. This happens because 1) Bayes-ToMoP$_{1}$ needs longer time to determine that the opponent is not using a BPR-based strategy; 2) some stochastic factors such as the random initialization of belief. This phenomenon is consistent with the slightly lower win rate of Bayes-ToMoP$_{1}$ than BPR$_s$ and deep BPR+ (Table \ref{table2}). 
The slight performance decrease against O$_{ns}$ is worthwhile since Bayes-ToMoP$_{1}$ performs much better than deep BPR+ against other two types of advanced opponents (O$_{ToMoP_{0}}$ and O$_{ToMoP_{0}-s}$) as shown in Table \ref{table2}. 
However, DRON only achieves the average rewards of 0.7 since it uses end-to-end trained response subnetworks, which cannot guarantee that each response is good enough against a particular type of opponent. 

\begin{figure}
   \subfloat[Thieves and hunters]{
  \begin{minipage}[t]{.47\linewidth}    
	\centerline{\includegraphics[height = 1in]{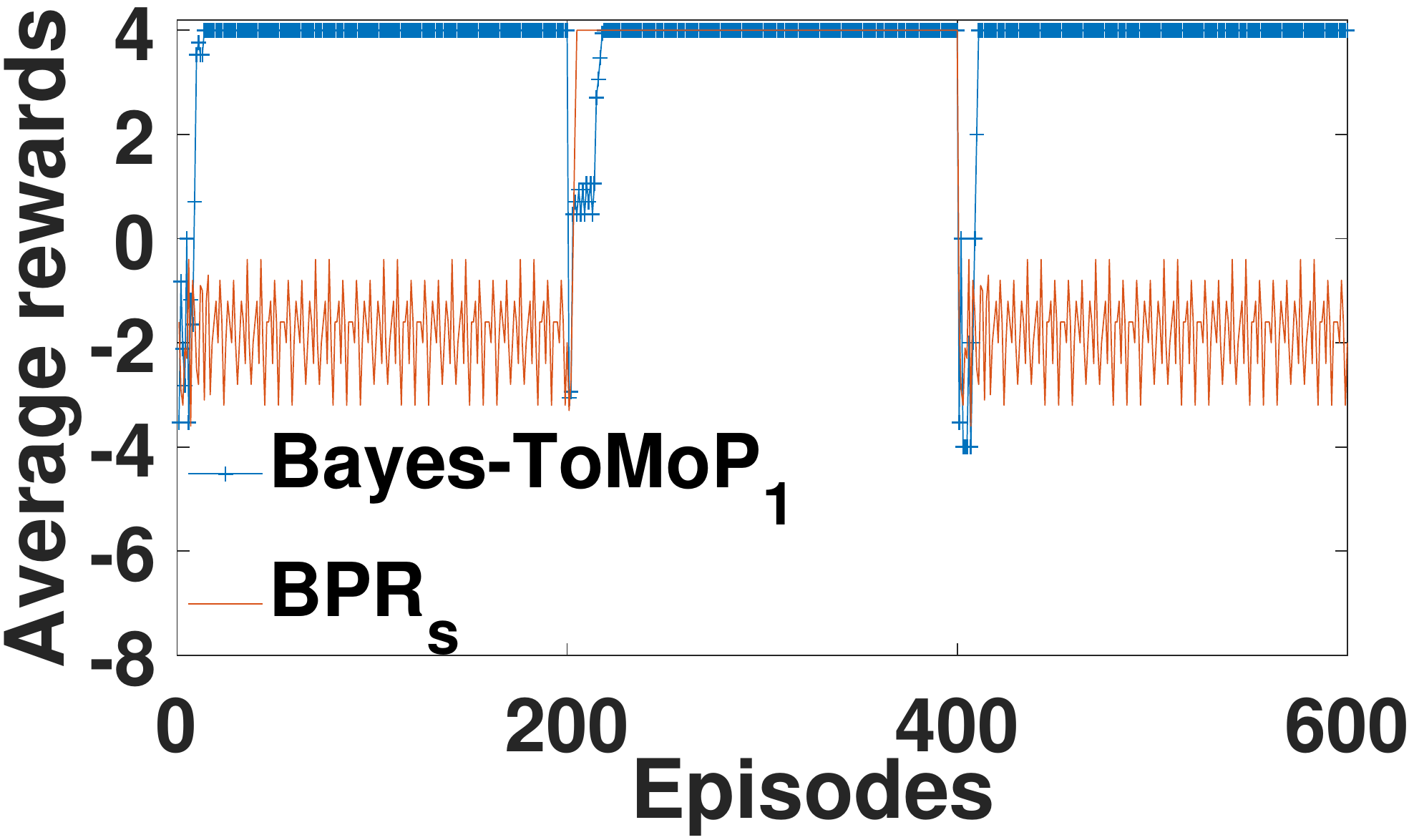}}	
  \end{minipage} 
 }
 \hfill
	\subfloat[Soccer (deep version)]{
   \begin{minipage}[t]{.47\linewidth}    
	\centerline{\includegraphics[height = 1in]{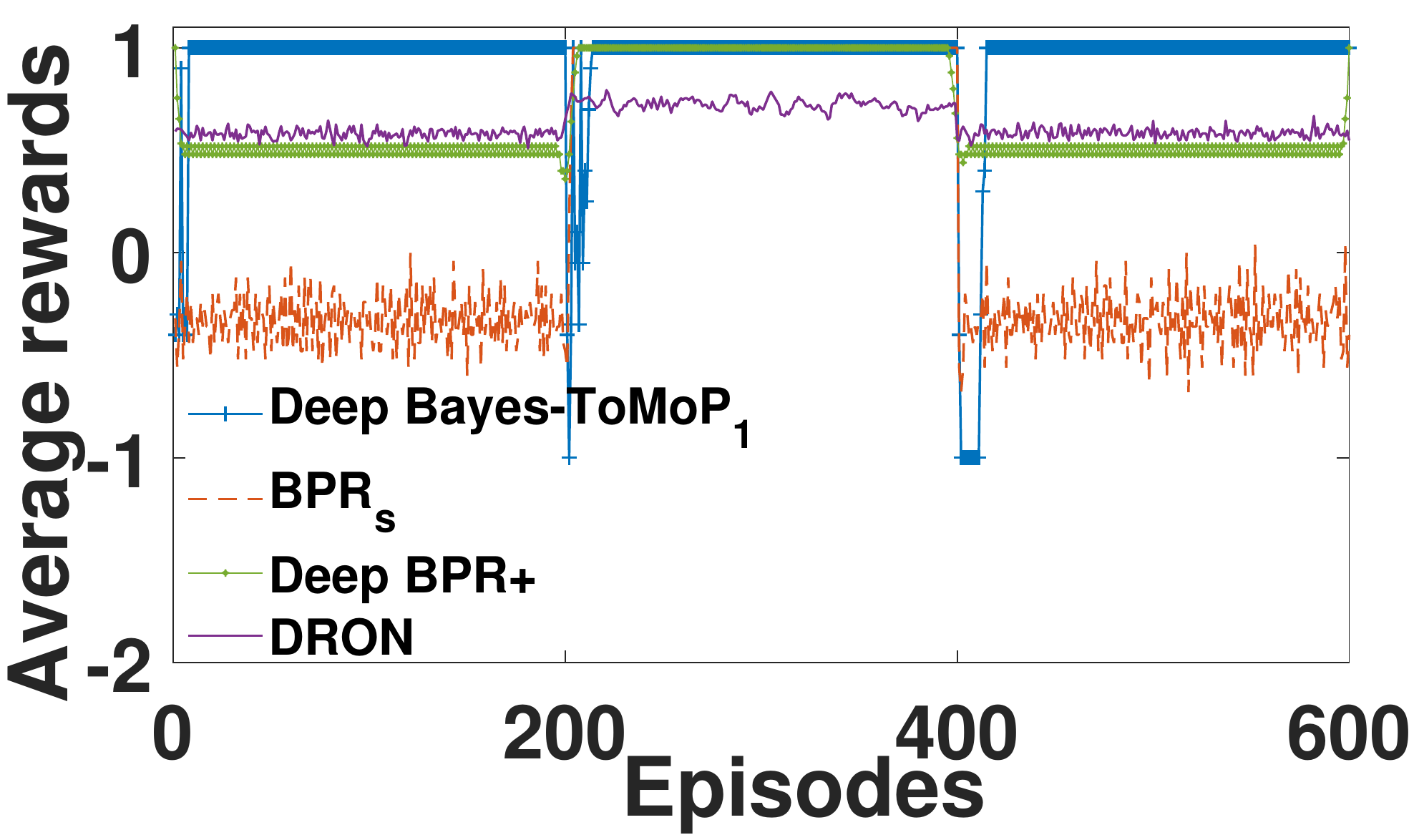}}	
  \end{minipage} 
   }
\caption{Against an opponent O$_{ToMoP_{0}-s}$ on different games.} \label{figure5}
\end{figure}

Finally, we consider the case of playing against O$_{ToMoP_{0}}$-$s$ to show the robustness of Bayes-ToMoP. Figure \ref{figure5} (a) shows that only Bayes-ToMoP$_{1}$ can quickly and accurately detect the strategies of both opponent O$_{ToMoP_{0}}$ and non-stationary opponent. In contrast, BPR$_s$ fails when its opponent's strategy switches to Bayes-ToMoP$_{0}$. A similar comparison can be found in soccer shown in Figure \ref{figure5} (b) in which both BPR$_s$ and deep BPR+ fail to detect and respond to O$_{ToMoP_{0}}$-$s$ opponent. Figure \ref{figure5} (b) also shows DRON performs poorly against both kinds of opponents due to the similar reason described above. Average win rates are summarized in Table \ref{table2} which are consistent with the results in Figure \ref{figure5} (b).

\subsection{New Opponent Detection and Learning}\label{sec5.4}

\begin{figure}
\subfloat[Thieves and hunters]{
    \begin{minipage}[t]{.47\linewidth}   
  	\centerline{\includegraphics[height = 1in]{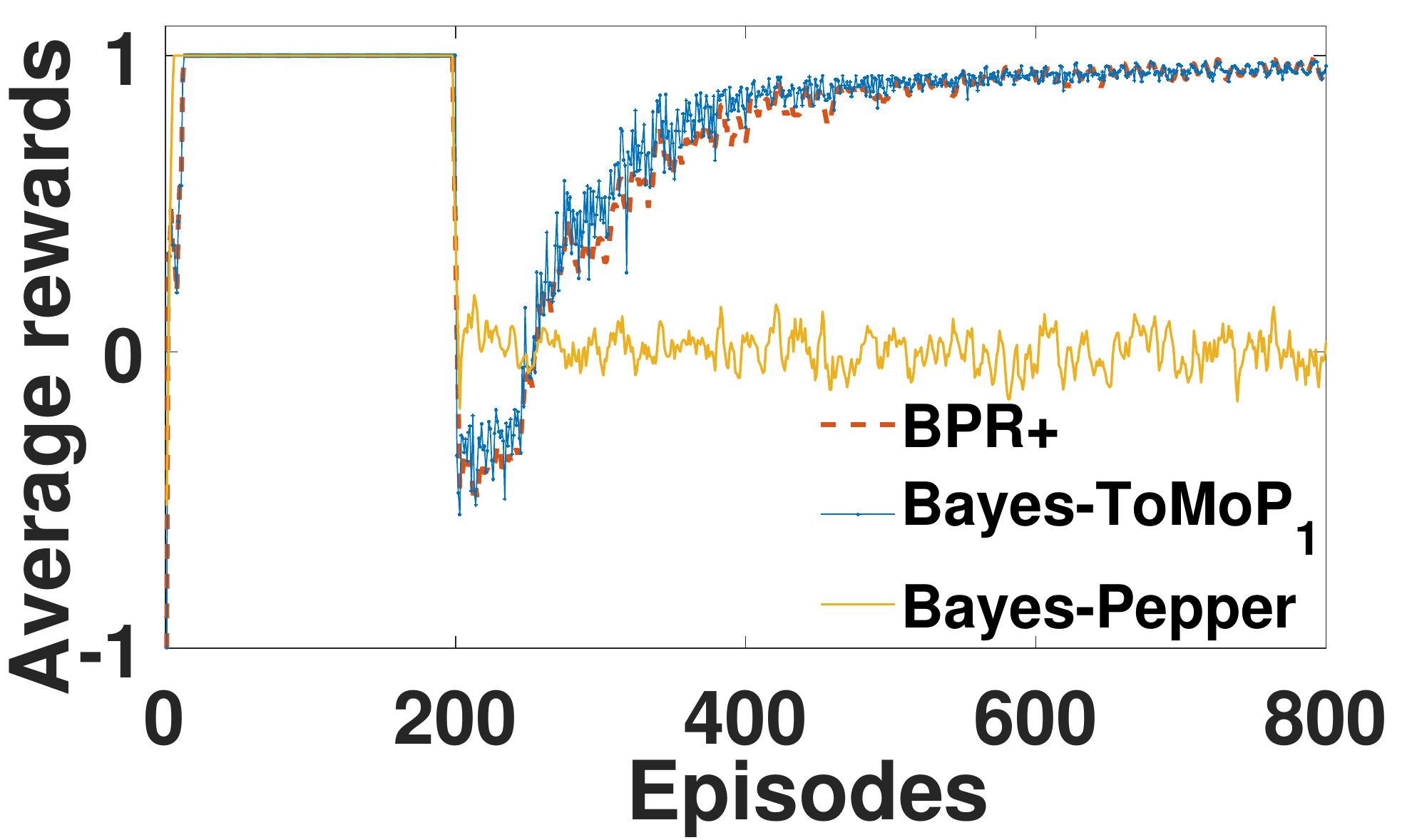}} 
  \end{minipage}
  }
  \hfill
  \subfloat[Soccer (deep version)]{
   \begin{minipage}[t]{.47\linewidth}    
	\centerline{\includegraphics[height = 1in]{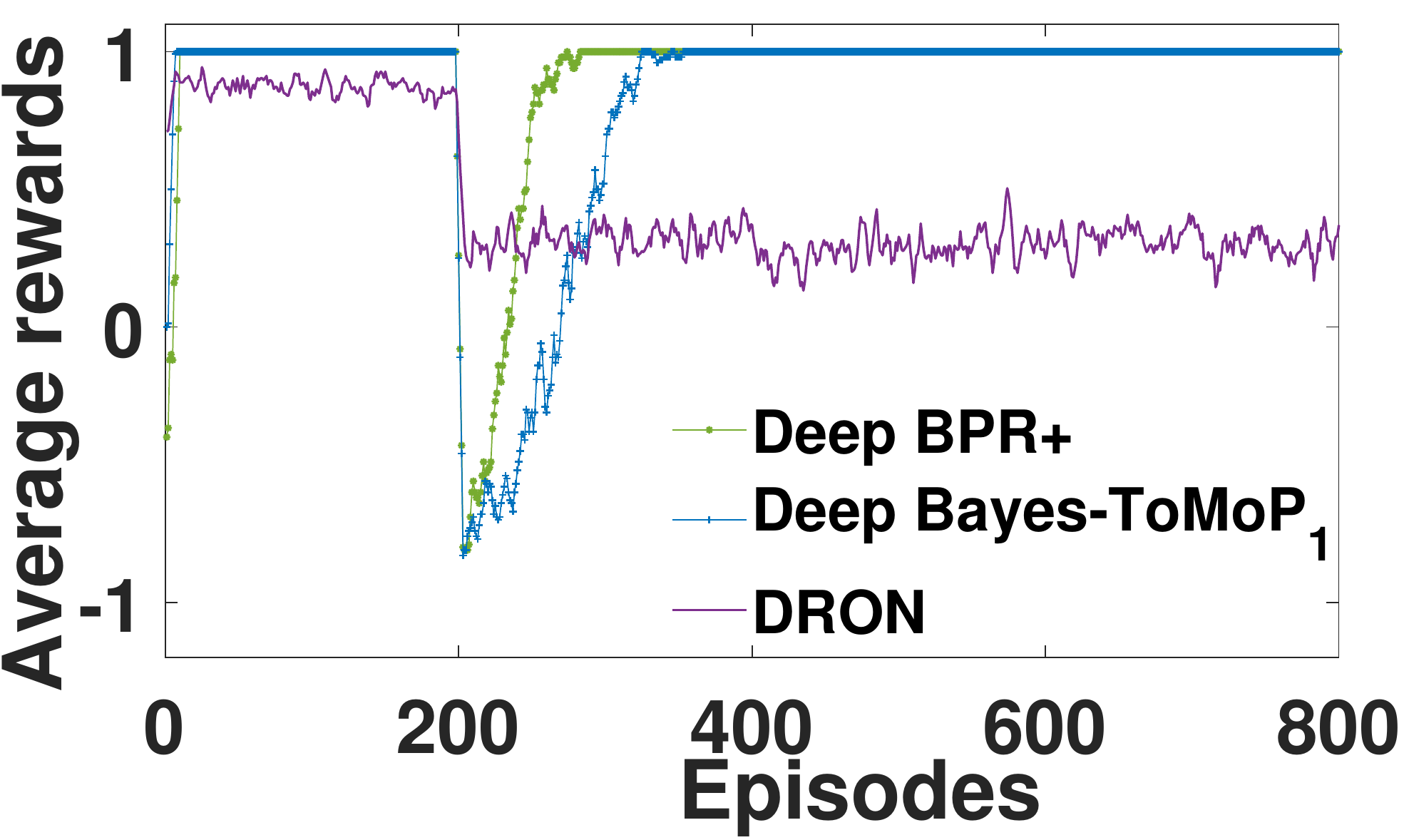}}
  \end{minipage}
  }
  \caption{Against an opponent using a new policy on two games.} \label{figure11}
\end{figure}

In this section, we evaluate Bayes-ToMoP$_{1}$ against an opponent who may use previously unknown strategies. We manually design a new opponent strategy in soccer (the new policy is different from the policies in the library at starting from a different grid and approaching the goal following a different direction) and thieves and hunters (different from the policies in the library at approaching the four destinations in a different order). The opponent stars with one of the known strategies and switches to the new one at the 200th episode. Figure \ref{figure11} shows the average rewards of different approaches on two games respectively. 

Figure \ref{figure11} (a) shows that when the opponent switches to an unknown strategy, both Bayes-ToMoP$_{1}$ and BPR+ can quickly detect this change, and finally learn an optimal policy. However, Bayes-Pepper fails. This is because Bayes-Pepper predicts the strategy of its opponent from a known policy library, which makes it fail to well respond to a previously unseen strategy. The learning curve of Bayes-ToMoP$_{1}$ is closer to BPR$+$ since both methods learn from scratch. Similar results can be found for deep Bayes-ToMoP$_1$ and BPR+ in Figure \ref{figure11} (b). DRON fails against the unknown opponent strategy due to the fact that the number of expert networks is fixed and thus unable to handle such a case. Deep BPR+ performs better than the other three methods since it uses policy distillation to transfer knowledge from similar previous policies to accelerate online learning, which can be readily applied to our Bayes-ToMoP to accelerate online learning as future work.


\section{Conclusion and Future Work} \label{sec6}
This paper presents a novel algorithm called Bayes-ToMoP to handle not only switching, non-stationary opponents and also more sophisticated ones (e.g., BPR-based). Bayes-ToMoP also enables an agent to learn a new optimal policy when encountering a previously unseen strategy. Theoretical guarantees are provided for the optimal detection of the opponent's strategies. Extensive simulations show Bayes-ToMoP outperforms the state-of-the-art approaches both in tabular and deep learning environments. 

Bayes-ToMoP can be seen as a generalized framework to reason and detect the policy change of an opponent, in which any recent deep RL algorithms can be incorporated to address large-scale state and action space problems. Currently we only use DQN to handle large-scale state space, while policy-based DRL (e.g., DDPG) can be used for problems with continuous actions. On the other hand, Bayes-ToMoP can only handle two-player games, and it is worthwhile investigating how to apply theory of mind in multiple agents’ scenarios to our Bayes-ToMoP framework as future work. Furthermore, how to accelerate the online new policy learning phase and higher orders of Bayes-ToMoP are worth investigating to handle more sophisticated opponents and apply to large scale, real scenarios.
\section*{Acknowledgments}
The work is supported by the National Natural Science Foundation of China (Grant Nos.: 61702362, U1836214), Special Program of Artificial Intelligence, Special Program of Artificial Intelligence and Special Program of Artificial Intelligence of Tianjin Municipal Science and Technology Commission (No.: 569 17ZXRGGX00150) and Huawei Noah's Ark Lab (Grant No.: YBN2018055043). 
%
%

\bibliographystyle{named}
\bibliography{ijcai19-bayes}

\section*{Supplementary Materials}
\begin{theorem} \label{theorem1}
\textbf{(Optimality on Strategy Detection)} If the opponent plays a strategy from the known policy library, Bayes-ToMoP can detect the strategy w.p.1 and selects an optimal response policy accordingly.
\end{theorem}


\begin{proof} \label{proof1}
Suppose the opponent strategy is $j \in \mathcal{J}$ at step $t$, the belief $\beta_{t}(j)$ is updated as follows: 
\begin{displaymath}
\beta_{t+1}(j)=\frac{P(\sigma_t|j, \pi)\beta_{t}(j)}{\sum_{j^{'}\in \mathcal{J}}P(\sigma_t|j^{'}, \pi)\beta_{t}(j^{'})}
\end{displaymath}
where the signal $\sigma_t$ is the average reward of policy $\pi$ against the opponent strategy over last $l$ episodes and can approximate the expected payoff.

Since Bayes-ToMoP already learned a best response against each opponent strategy in the offline phase, given the expected payoff against an opponent’s strategy, there always exists a corresponding best-response policy $\pi \in \Pi$, that makes the inequality $P(\sigma_t|j,\pi)\geq P(\sigma_t|j^{'},\pi)$ establish for all $\forall j^{'} \in \mathcal{J}$, i.e., the probability of receiving the expected payoff is larger than or equal to others. The equality holds when one policy can beat two or more types of opponent strategies. Besides, since $\beta^{t}(j)$ is bounded ($0\leq \beta_{t}(j) \leq 1$) and monotonically increasing ($\beta^{t+1}(j)-\beta^{t}(j)>0$) which is deduced by the above-mentioned equation, based on the monotone convergence theorem \cite{rudin1964principles}, we can easily know the limit of sequence $\beta^{t}(j)$ exists. Thus, if we limit the two sides of the above Equation, the following equation establishes,

\[\begin{array}{l}
{\rm{P}}({\sigma _t}|j,\pi ) = \sum\limits_{j' \in \mathcal{J}} {{\rm{P}}({\sigma _t}|j',\pi )} {\beta _t}(j'),\\
if{f_{}}\quad\forall j' \ne j,{\beta _t}(j) = 1,{\beta _t}(j') = 0
\end{array}\]
So that Bayes-ToMoP can detect the strategy w.p.1 and selects an optimal response policy accordingly.
\end{proof}

As guaranteed by Theorem \ref{theorem1}, Bayes-ToMoP behaves optimally when the opponent uses a strategy from the known policy library. If the opponent is using a previously unseen strategy, following the new opponent strategy detection heuristic, this phenomenon can be exactly detected when the winning rate of Bayes-ToMoP during a fixed length of episodes is lower than the accepted threshold. Then Bayes-ToMoP begins to learn an optimal response policy. 

\section*{Network Architecture}
All experiments use the same parameter settings: $c_{1} = 0.3, \lambda = 0.7, \delta = 0.7$ (experimentally selected). For deep Bayes-ToMoP$_{1}$, we consider the DNN input which consists of different dimensions of environment information: for example, states in soccer includes coordinates of two agents and the ball possession. DQN we used has two fully-connected hidden layers both with 20 hidden units, the output layer is a fully-connected linear layer with a single output for each valid action. We train DQN each step with mini-batches of size 32 randomly sampled from a replay buffer of one million transitions. Parameters are copies to the target network every 500 episodes. The learning rate of DQN is $10^{-3}$ and the discount factor is 0.9. All results are averaged over 1000 runs.
\begin{figure}[H]
\centering
\subfloat[RPS]{
\begin{minipage}[t]{.47\linewidth} 
	\centerline{\includegraphics[height = 1in]{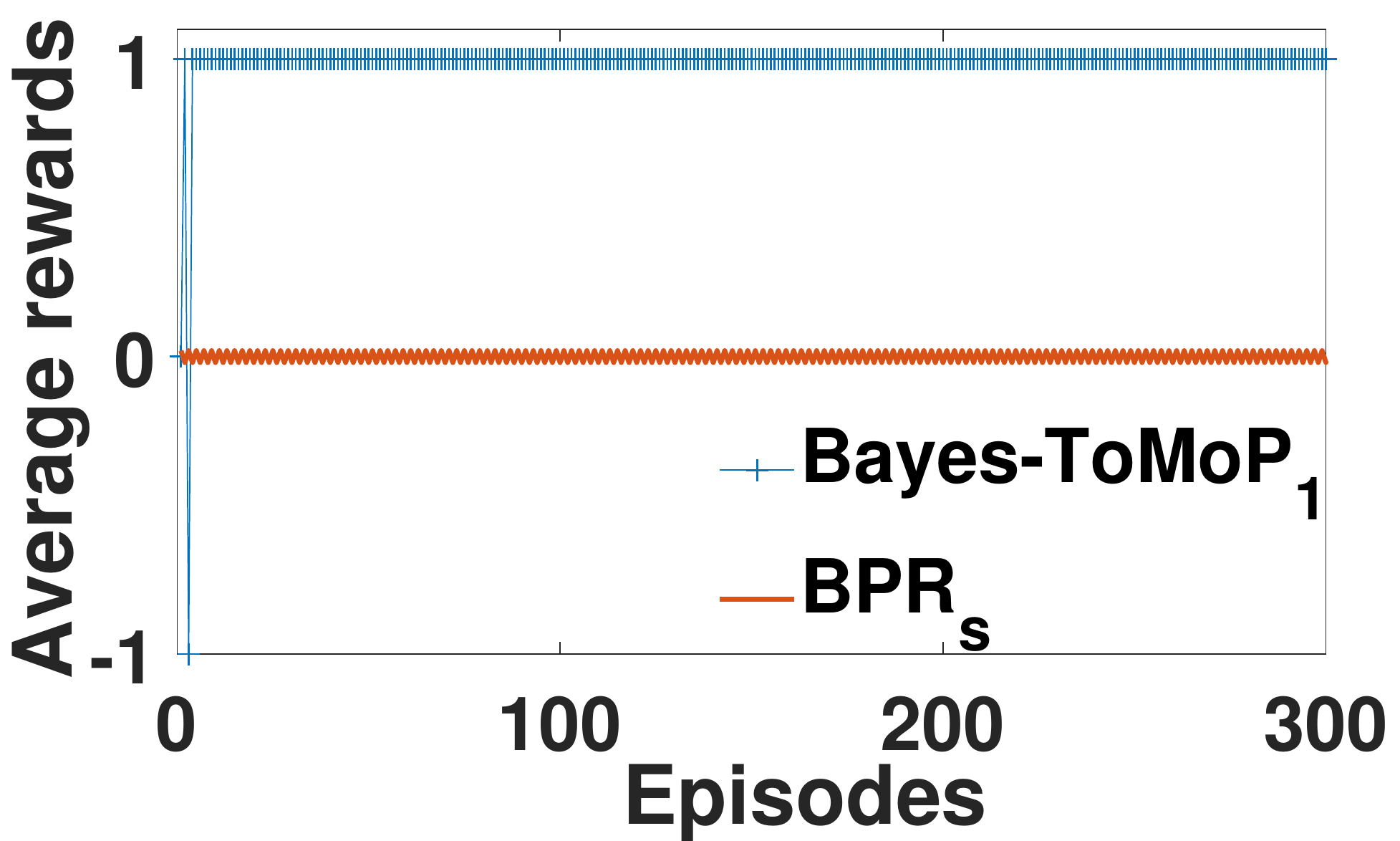}}	
	\end{minipage} 
	}
	\subfloat[Tabular soccer]{
	\hfill
   \begin{minipage}[t]{.47\linewidth}    
	\centerline{\includegraphics[height = 1in]{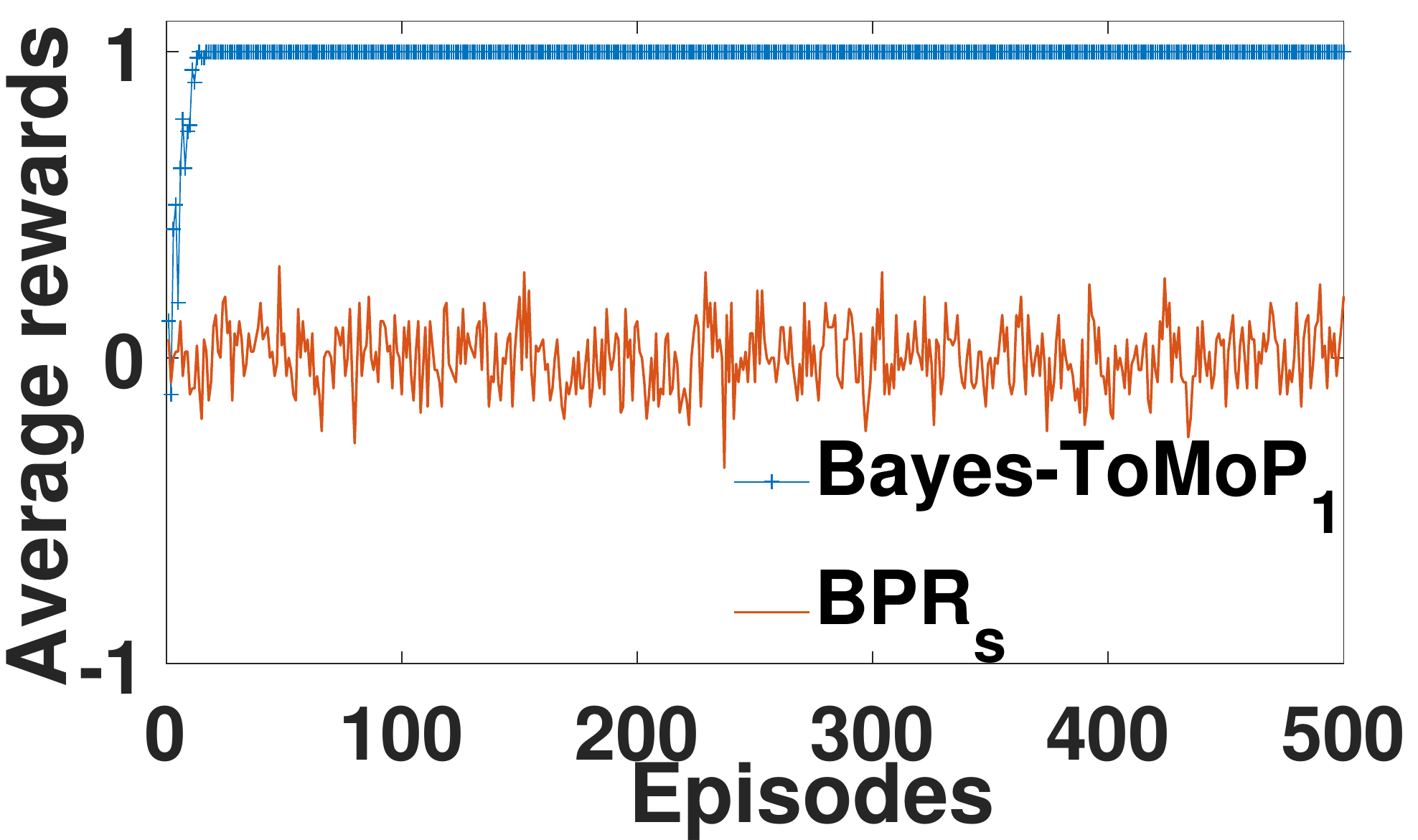}}	
  \end{minipage} 
   }
   \caption{Different approaches against an opponent O$_{ToMoP_{0}}$ on different games.} \label{s1}
  \end{figure}

\section*{Performance against Different Opponents}
\begin{figure}[H]
\subfloat[RPS]{
	\hfill
   \begin{minipage}[t]{.47\linewidth}    
	\centerline{\includegraphics[height = 1in]{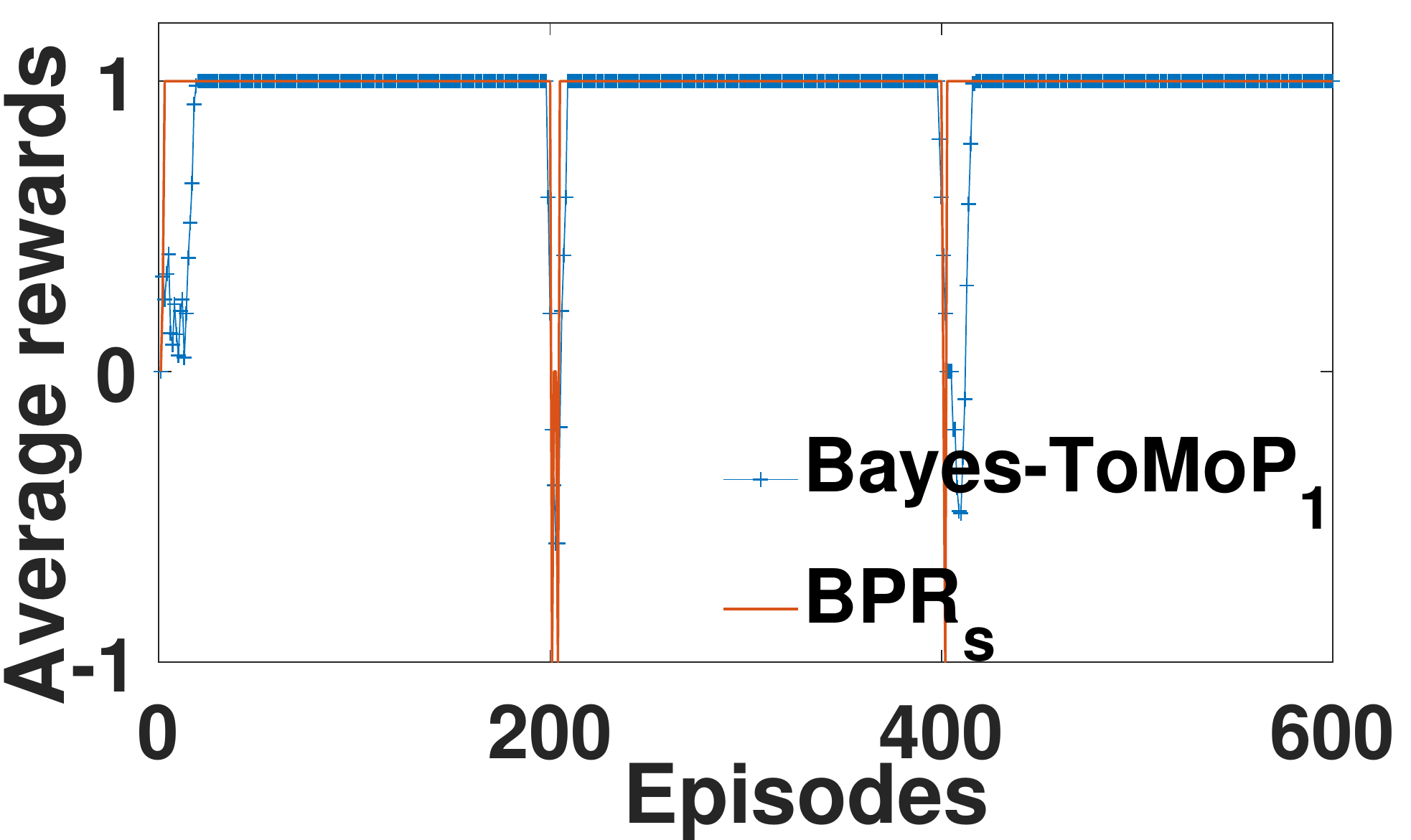}}	
  \end{minipage} 
   }
   \hfill
	\subfloat[Tabular soccer]{
   \begin{minipage}[t]{.47\linewidth}    
	\centerline{\includegraphics[height = 1in]{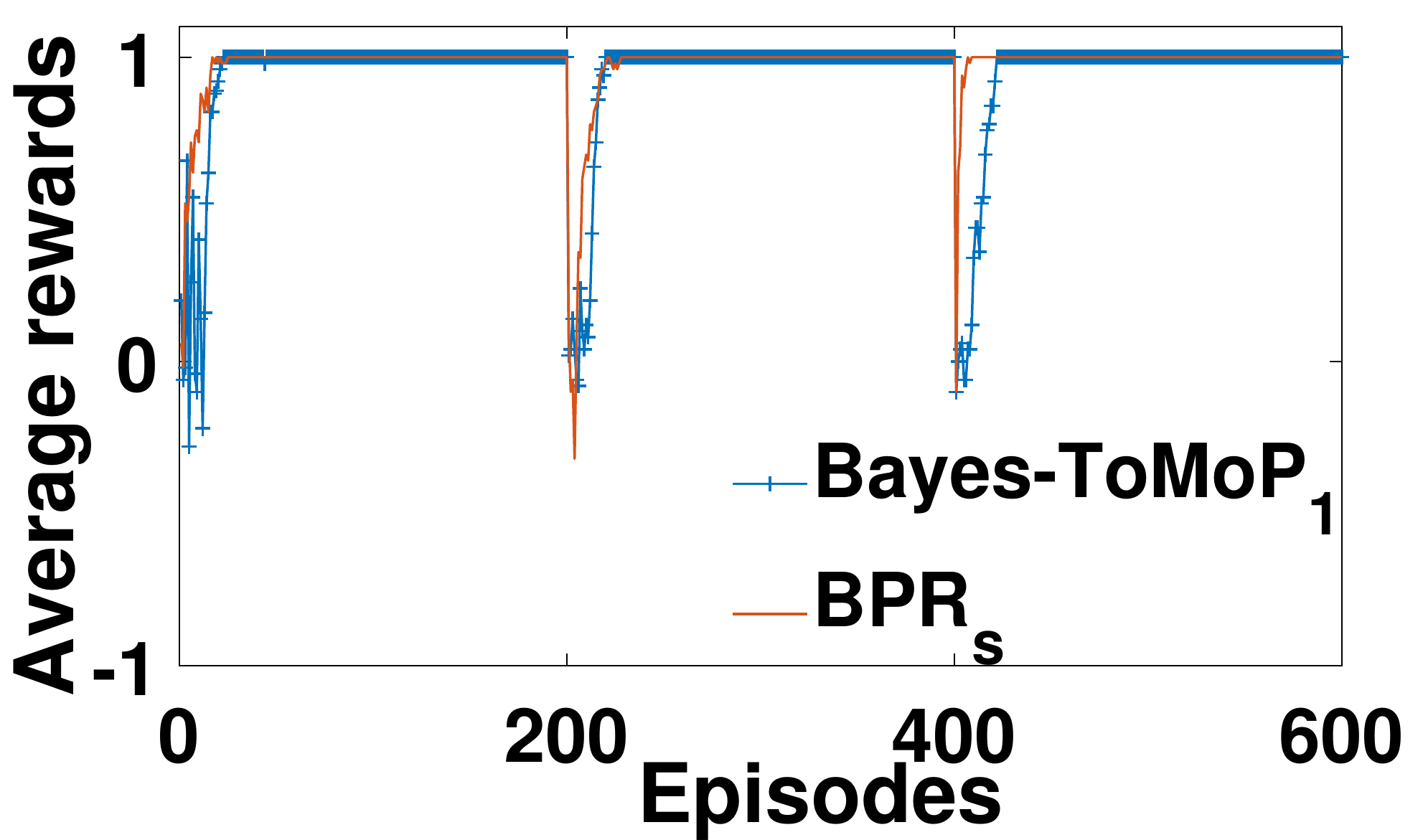}}	
  \end{minipage} 
   }
   \caption{Different approaches against an opponent O$_{ns}$ on different games.}\label{s2}
  \end{figure}

\begin{figure}
\subfloat[RPS]{
	\hfill
   \begin{minipage}[t]{.47\linewidth}    
	\centerline{\includegraphics[height = 1in]{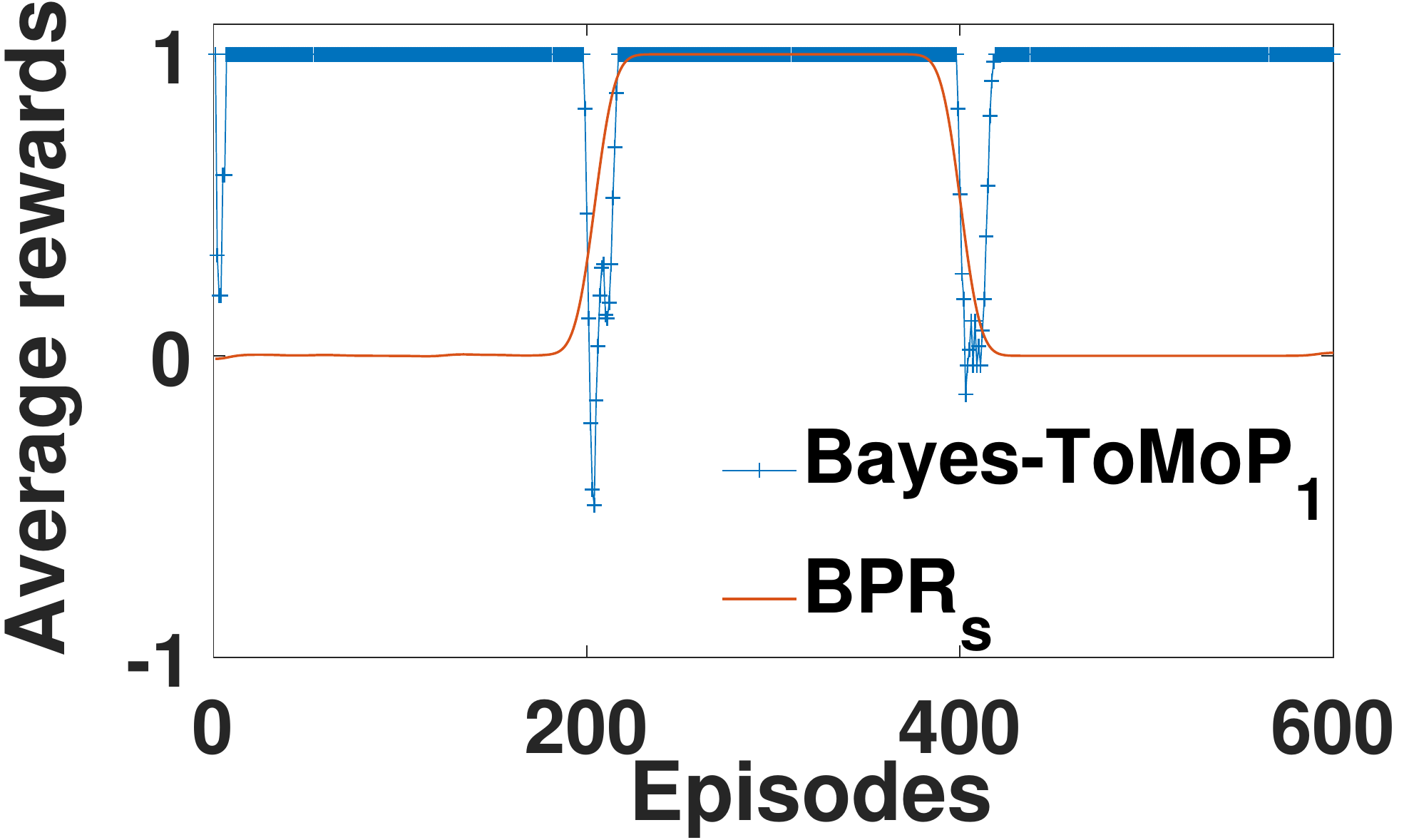}}	
  \end{minipage} 
   }
	\subfloat[Tabular soccer]{
	\hfill
   \begin{minipage}[t]{.47\linewidth}    
	\centerline{\includegraphics[height = 1in]{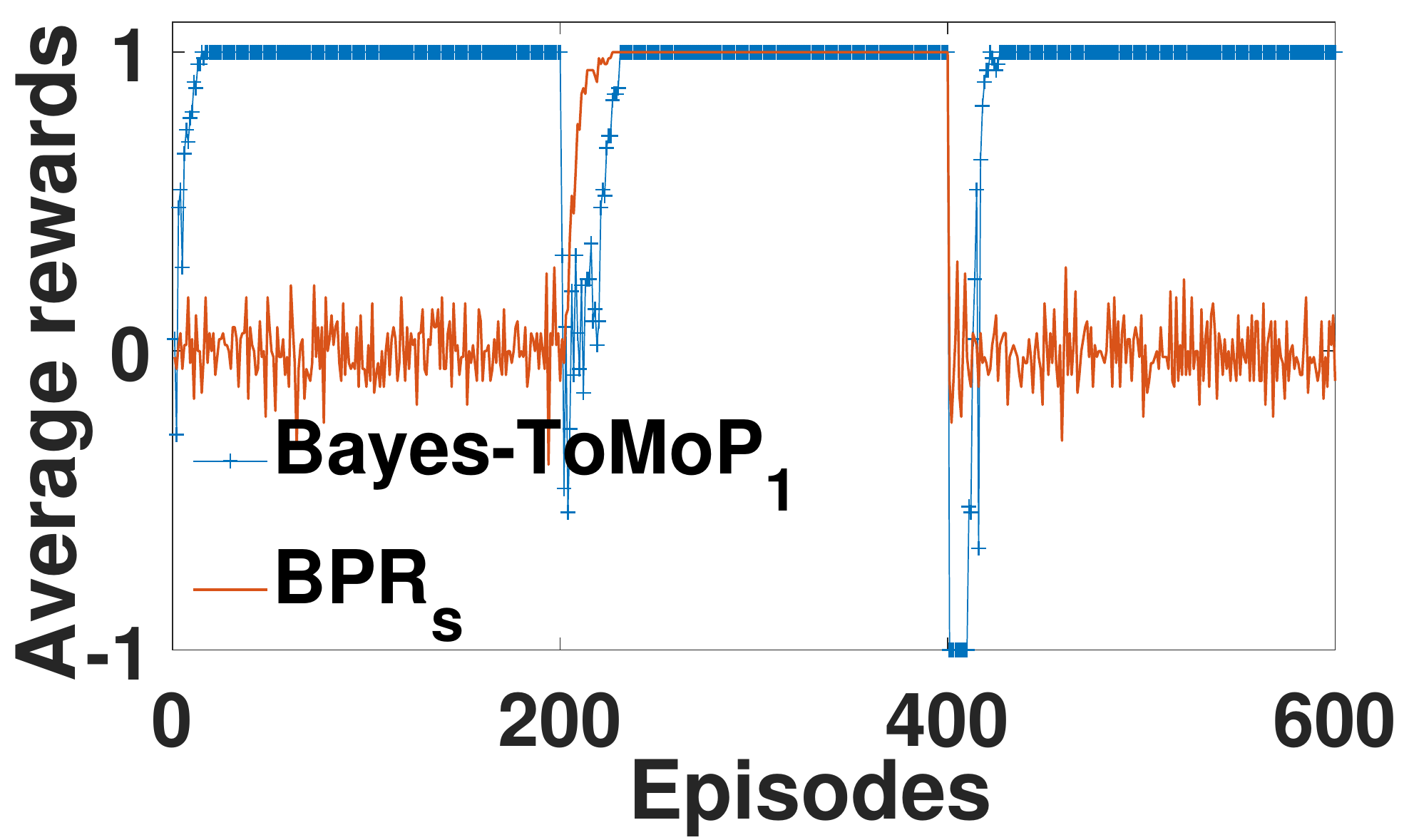}}	
  \end{minipage} 
   }
   \caption{Different approaches against an opponent O$_{ToMoP_{0}-s}$ on different games.}\label{s3}
  \end{figure}

Rock-paper-scissors (RPS) \cite{J1928Zur,Shoham2009Multiagent} is a two-player stateless game in which two players simultaneously choose one of the three possible actions `rock' (R), `paper' (P), or `scissors' (S). If both choose the same action, the game ends in a tie. Otherwise, the player who chooses R wins against the one that chooses S, S wins against P, and P wins against R. Results on RPS and soccer games are shown as follows (Figure \ref{s1}, \ref{s2}, \ref{s3}).

\section*{Bayes-ToMoP$_{1}$ under Self-play}

\begin{figure}[H]
\centerline{\includegraphics[height = 1in]{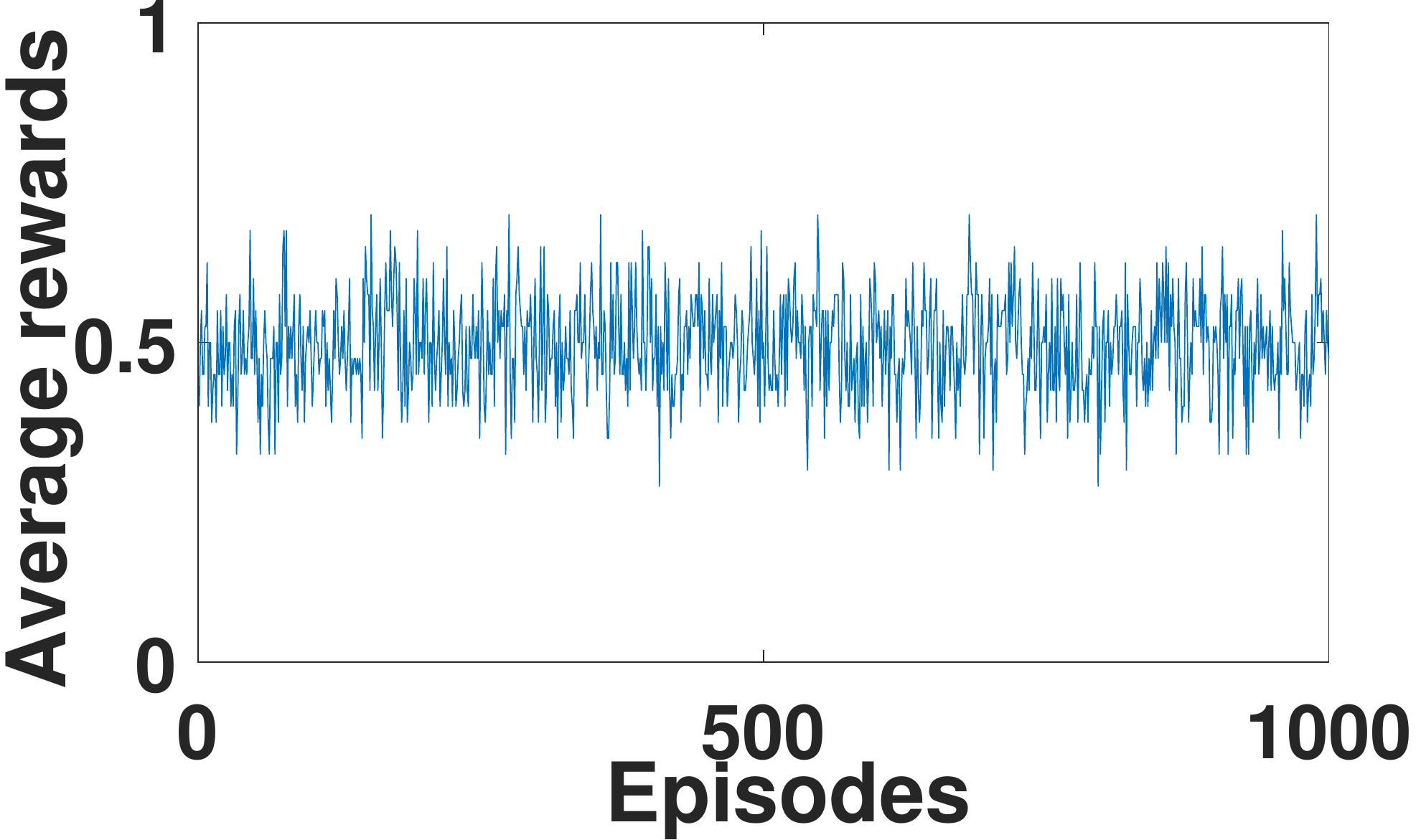}}
	\caption{Bayes-ToMoP$_{1}$ under self-play.} \label{figuretom1}
\end{figure}

Results (Figure \ref{figuretom1}) are the same as Bayes-ToMoP$_{0}$ under self-play due to the similar reasons, thus it is worth investigating higher order of Bayes-ToMoP to handle more kinds of opponents.

\section*{The Influence of Key Parameters}\label{sec5.5}

In this section, we analyze the influence of key parameters on the performance of Bayes-ToMoP, e.g., the threshold $\delta$ and the memory length $l$. 

\begin{figure}[H]
\hfill
   \begin{minipage}[t]{.48\linewidth}    
	\centerline{\includegraphics[height = 1in]{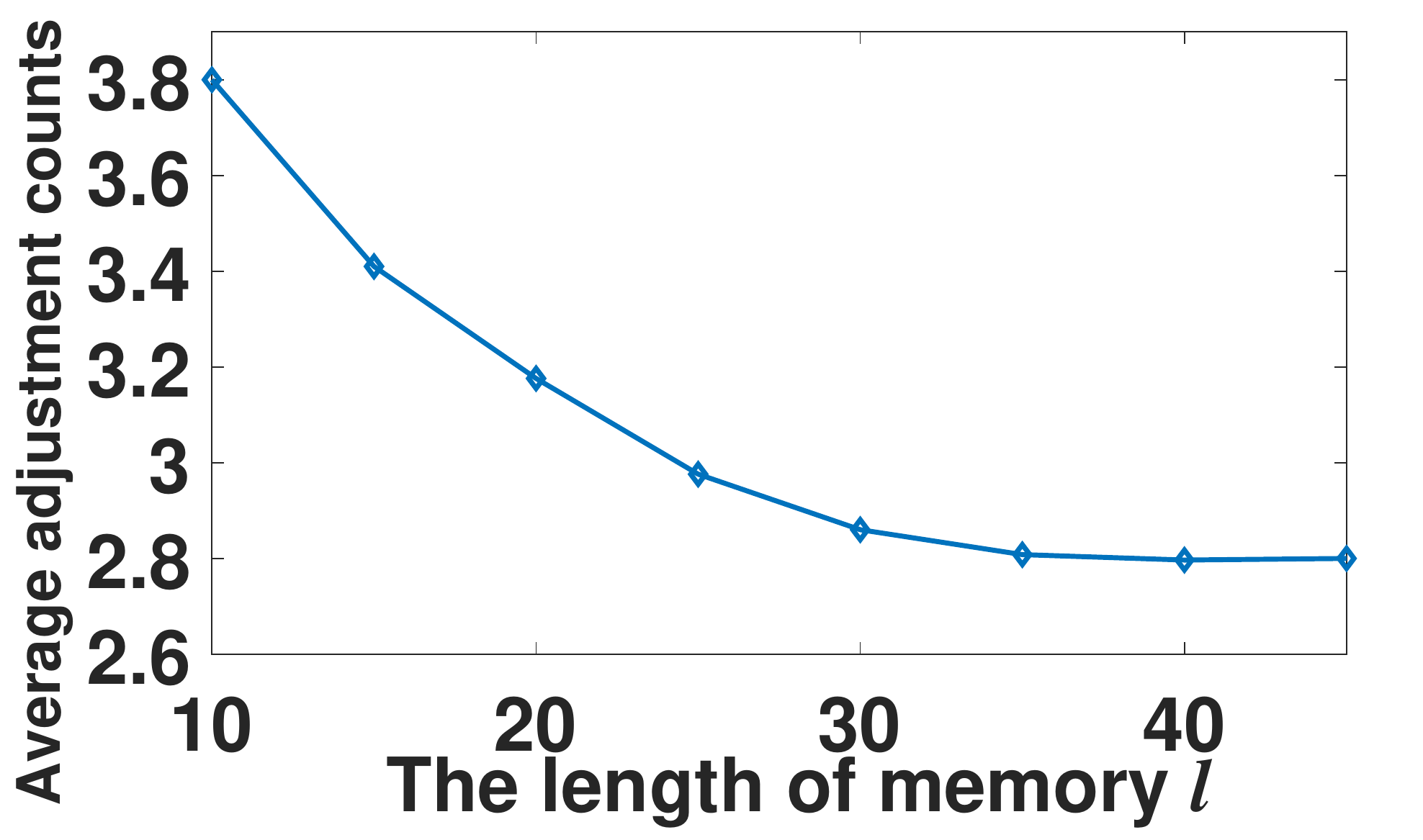}}
	\caption{The impact of memory length $l$.} \label{figure7}
  \end{minipage}
  \hfill
   \begin{minipage}[t]{.48\linewidth}    
	\centerline{\includegraphics[height = 0.95in,width=1.75in]{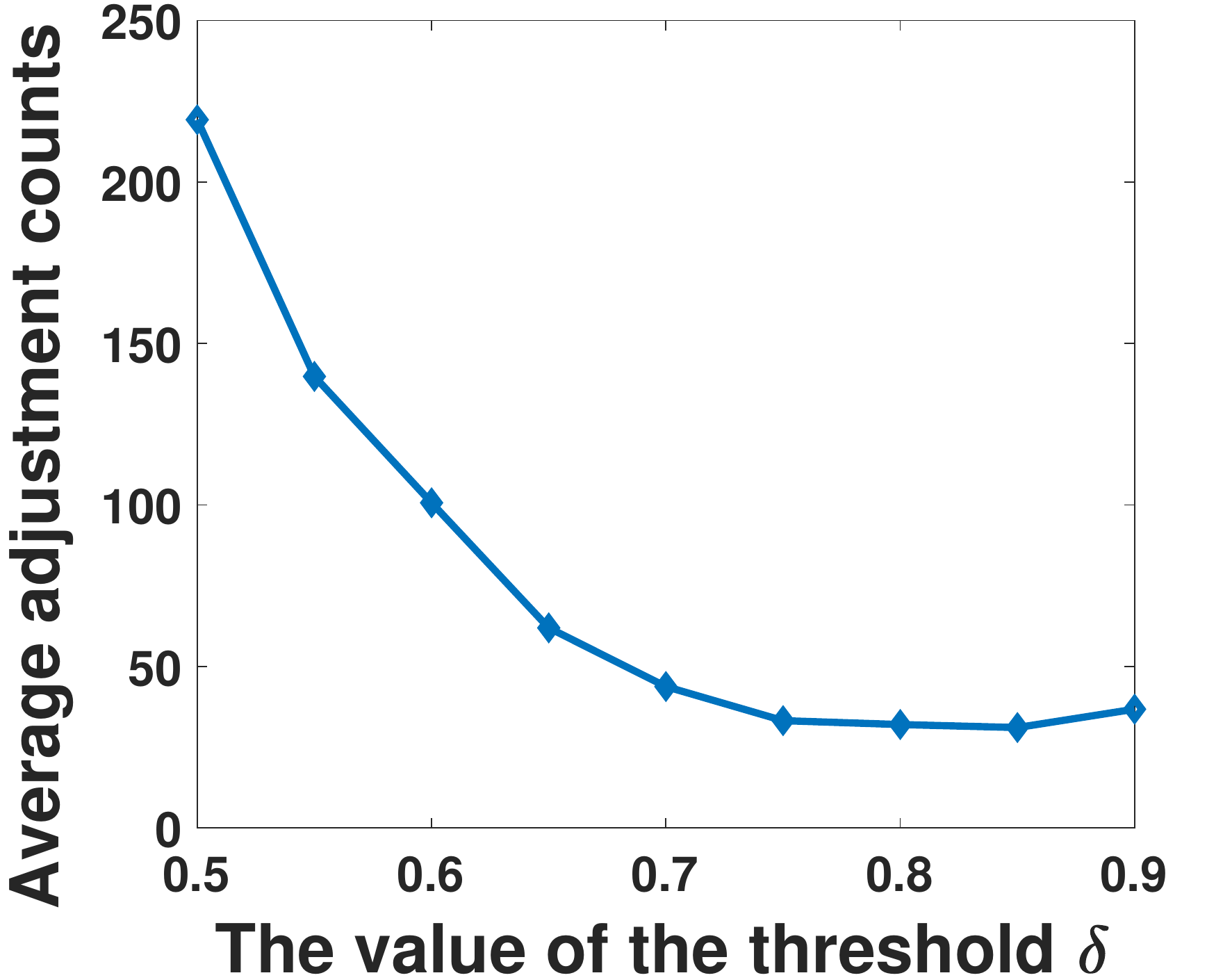}}
	\caption{The impact of threshold $\delta$.} \label{figure10}
  \end{minipage}
\end{figure}

Figure \ref{figure7} depicts the impact of the memory length $l$ on the average adjustment counts of Bayes-ToMoP$_{1}$ before taking advantage of O$_{ToMoP_{0}}$. We observe a diminishing return phenomenon: the average adjustment counts decrease quickly as the initial increase in the memory length, but quickly render additional performance gains marginal. The average adjustment counts stabilize around 2.8 when $l>35$. We hypothesize that it is because the dynamic changes of the winning rate over a relatively small length of memory may be caused by noise thus resulting in inaccurate opponent type detection. As the increase of the memory length, the judgment about the opponent's types is more precise. However, as the memory length exceeds a certain threshold, the winning rate estimation is already accurate enough and thus the advantage of further increasing the memory length diminishes. 

Finally, the influence of threshold $\delta$ against opponent O$_{ns}$ is shown in Figure \ref{figure10}. We note that the average adjustment counts decrease as $\delta$ increases, but the decrease degree gradually stabilizes when $\delta$ is larger than 0.7. With the increase of the value of $\delta$, the winning rate decreases to $\delta$ more quickly when the opponent switches its policy. Thus, Bayes-ToMoP$_1$ detects the switching of the opponent's strategies more quickly. Similar with the results in Figure \ref{figure7}, as the value of $\delta$ exceeds a certain threshold, the advantage based on this heuristic diminishes.
\end{document}